\documentclass[11pt]{article}
\usepackage[T1]{fontenc}

\usepackage{subcaption}
\usepackage{hyperref}
\usepackage{amsmath, amssymb, amsthm, float, braket, mathrsfs}
\usepackage{bbold}
\usepackage{bm}
\usepackage{tikz}
\usepackage{tikz-3dplot}
\usepackage{tikz-cd}
\usepackage[utf8]{inputenc}
\usepackage[english]{babel}
\usepackage{imakeidx}

\DeclareMathOperator{\Tr}{Tr}

\DeclareMathOperator{\SO}{SO}
\DeclareMathOperator{\SU}{SU}
\DeclareMathOperator{\su}{\mathfrak{su}}

\DeclareMathOperator{\SPD}{SPD}
\DeclareMathOperator{\Id}{Id}

\newcommand{\be}{\begin{equation}}
\newcommand{\ee}{\end{equation}}
\newcommand{\bea}{\begin{eqnarray}}
\newcommand{\eea}{\end{eqnarray}}

\makeatletter

\usepackage{babel}
\numberwithin{equation}{section}
\begin{document}
\immediate\write16{<<WARNING: LINEDRAW macros work with emTeX-dvivers
                    and other drivers supporting emTeX \special's
                    (dviscr, dvihplj, dvidot, dvips, dviwin, etc.) >>}

\title{\textbf{Compressibility of dense nuclear matter in the $\rho$-meson variant of the Skyrme model}}

\author{\large Miguel Huidobro\textsuperscript{1}\,, Paul Leask\textsuperscript{2,3}\footnote{palea@kth.se}\,, Carlos Naya\textsuperscript{4,5}\,,\: and Andrzej Wereszczynski\textsuperscript{5,6,7} \\\\
\footnotesize\textsuperscript{1}\textit{Departamento de F\'isica de Part\'iculas, Universidad de Santiago de Compostela and} \\
\footnotesize\textit{Instituto Galego de F\'isica de Altas Enerxias (IGFAE), Santiago de Compostela, E-15782, Spain} \\
\footnotesize\textsuperscript{2}\textit{School of Mathematics, University of Leeds, Leeds LS2 9JT, UK} \\
\footnotesize\textsuperscript{3}\textit{Department of Physics, KTH Royal Institute of Technology, 10691 Stockholm, Sweden} \\
\footnotesize\textsuperscript{4}\textit{Universidad de Alcal\'a, Departamento de F\'isica y Matem\'aticas, 28805 Alcalá de Henares, Spain} \\
\footnotesize\textsuperscript{5}\textit{Institute of Physics, Jagiellonian University, Lojasiewicza 11, Krak\'{o}w, Poland} \\
\footnotesize\textsuperscript{6}\textit{Department of Applied Mathematics, University of Salamanca, Casas del Parque 2}, \\
\footnotesize\textit{and Institute of Fundamental Physics and Mathematics, University of Salamanca}, \\
\footnotesize\textit{Plaza de la Merced 1, 37008 - Salamanca, Spain}, \\
\footnotesize\textsuperscript{7}\textit{International Institute for Sustainability with Knotted Chiral Meta Matter (WPI-SKCM2)}, \\  \footnotesize\textit{Hiroshima University, Higashi-Hiroshima, Hiroshima 739-8526, Japan}, 
\\}

\maketitle

\begin{abstract}
We show that coupling the $\SU(2)$-valued Skyrme field to the $\rho$-meson solves the long-standing issue of (in)compressibility in the solitonic Skyrme model. 
Even by including only one $\rho\pi$ interaction term, motivated by a holographic-like reduction of Yang-Mills action by Sutcliffe, reduces the compression modulus from $K_0 \simeq 1080$ MeV, in the massive Skyrme model, to $K_0\simeq 351$ MeV.
\end{abstract}


\section{Introduction}

Although the fundamental theory of strong interactions is known, a derivation of properties of baryons, atomic nuclei and (infinite) nuclear matter is still one of the main challenges of the contemporary theoretical physics.
This is due to the well known fact that quantum chromodynamics (QCD) is a strongly interacting quantum theory in the low energy limit, where fundamental degrees of freedom, i.e., quarks and gluons, are confined into non-perturbative objects.
Consequently, the power-full perturbative expansion does not apply in this regime and we are left with lattice QCD or with the effective field theory (EFT) approach. 
 
The solitonic Skyrme model \cite{Skyrme_1961} (for a recent review see, e.g., \cite{Manton_2022}) is a very attractive example of an EFT which is not only well anchored in the underlying fundamental theory but, if contrasted with other EFT, has an extremely limited number of free parameters and therefore potentially possesses a huge predictive power. 
Skyrme's original theory consists of an effective mesonic Lagrangian involving only the lightest Goldstone bosons of QCD, the pions, with baryons emerging as stable topological solitons.
The theory has $N_f=2$ flavours of quarks (u,d) and the pion fields are encoded in the $\SU(N_f=2)$-valued Skyrme field.

The predicative power of the model has recently been proven in many cases, elevating the Skyrme model from a qualitative framework to a theory which allows for very accurate computations.
The main steps are the {\it vibrational quantization} \cite{Halcrow_2016}, or in generality inclusion of vibrational, massive deformations of Skyrmions \cite{Gudnason_2018}, and inclusion of additional terms, as e.g., the {\it sextic term} \cite{Adam_2010} or additional degrees of freedom, i.e., {\it vector mesons} \cite{Sutcliffe:2010et}. 

All that led to a very spectacular progress in the application of the Skyrme model to the description of atomic nuclei.
Let us only mention the computation of vibrational-rotational bands of light nuclei, as $^{7}$Li \cite{Halcrow_2016}, $^{12}$C \cite{Lau_2014, Adam_2024} and $^{16}$O \cite{Halcrow_2017, Manton_2019}, observation of $\alpha$-particle clustering in atomic nuclei, computation of nuclear forces \cite{Halcrow:2020gbm} and resolution of the too large binding energies problem (see also \cite{Adam_2010}, \cite{Sutcliffe:2010et}, \cite{Gillard_2015}, \cite{Naya_2018}, \cite{Gudnason_2020}).
There has also been a significant progress in description of nuclear matter and neutron stars with very good agreement with astrophysical observables, such as the mass-radius curve \cite{Naya_2019, Adam_2020, Adam_2022, Wereszczynski_2022}.
In particular, taking advantage of the new approach to determining phases of nuclear matter within the Skyrme model \cite{Leask_2023}, an equation of state covering the nuclear matter at all densities has been obtained \cite{Leask_2024}.  
 
However, these unquestionable successes all suffer from the last great issue of the Skyrme model which is the so-called compression modulus problem.
This was recently remedied by Harland \textit{et al}. \cite{Leask_Harland_2024} by investigating crystalline configurations in the Adkins--Nappi $\omega$-meson variant of the Skyrme model \cite{Nappi_1984}.
Therein they combined the method for determining crystals in the massive Skyrme model \cite{Leask_2023} with the method developed to obtain energy minimizers in the $\omega$-Skyrme model \cite{Gudnason_2020}.
In this letter we follow a similar approach where we study light nuclei and crystalline configurations in the massive Skyrme model coupled to the vector $\rho$-meson.
Using the ground state crystal in this theory, we are able to address the compression modulus problem.

\section{Compression modulus problem}
\label{problem}
 
The (symmetric) energy $E(n_B)$ of isospin symmetric nuclear matter can be approximated about the nuclear saturation density $n_0$ by use of a power series expansion in the baryon density $n_B$, that is \cite{Garg_2018}
\begin{equation}
    E(n_B) = E_0 + \frac{1}{2} K_0 \frac{(n_B-n_0)^2}{9n_0^2} + \mathcal{O} \left( (n_B-n_0)^3 \right),
\label{eq: Symmetric energy expansion}
\end{equation}
The first multipole in this expansion is the energy per baryon $E_0$ of symmetric nuclear matter at the saturation density.
There is no linear term since symmetric nuclear matter reaches a minimum of the energy at saturation.
The next multipole, $K_0$ is the compression modulus.
This is a fundamental quantity in nuclear physics which describes a resistance (stiffness) of nuclear matter under external force (pressure) at the saturation point. 

In practice, the compression modulus is not directly measured, but it may be extracted from the Isoscalar Giant Monopole Resonance (ISGMR) frequency, $\omega_M$.
This resonance is a collective excitation of the nucleus, in which both protons and neutrons vibrate spherically in phase.
It is measured through the low-momentum transfer in inelastic scattering collisions between isoscalar particles (like $\alpha$ particles or deuterons) and medium-heavy nuclei, $B \geq 90$.
The frequency of this resonance for a given nucleus may be related to its compression modulus $K$,
\begin{equation}
    \omega^2_M=\frac{K}{m_NR^2},
\end{equation}
where $R$ is the radius of the nucleus and $m_N$ is the mass of the nucleon. The value of this frequency was experimentally determined for different nuclei, such that the following dependence on the baryon number was found, $\hbar \omega_M \sim 80 B^{-1/3}$ MeV \cite{Colo_2004}.
Then, from the well-known relation between the radius of a nucleus and its baryon number, $R \approx 1.25 B^{1/3}$ fm, we find the $B$-independent value for the compression modulus \cite{Blaizot_1980},
\begin{equation}
K=K_0 \sim 240 \pm 20 \mbox{MeV}. 
\end{equation}
Further numerical simulations from theoretical approaches yield a range of
values for K which confirmed this value. 

To compute the compression modulus in the Skyrme model we use an equivalent definition from the Taylor expansion
\begin{equation}
 \left. K_0=9n_0^2 \frac{\partial^2 E}{\partial n_B^2} \right|_{n_0} =  \left.  9 V^2 \frac{\partial^2 E}{\partial V^2} \right|_{V_0}
\end{equation}
applied to Skyrme crystal i.e., a periodic classical solution of the Skyrme model representing infinite symmetric nuclear matter.
Here $V_0=B/n_0$ is the volume of the unit crystal cell at saturation, i.e. the volume at which the energy is minimal.
If the ground state crystalline configuration has cubic symmetry, then there is a corresponding energy minimizing length $L_0$ associated to $V_0=L_0^3$.

It is a matter of fact that the compression modulus in the Skyrme theory is a few times larger than the experimental value.
For different versions of the model and different crystal solutions its value ranges from $\sim 900$ MeV and $\sim 1350$ MeV for the massless and massive Skyrme to even higher values if the sextic term is added (with a not too high value of its coupling constant).
This gets an improvement if one uses the recently found mutli-wall solution which is the global crystal minimum provided the unit cell has $B_{\textup{cell}}=4$ baryon charge.
However, the value is still more than three times above the accepted value \cite{Leask_2024}.

The appearance of a too large value of the compression modulus can be easily understood in the minimal massless $\mathcal{L}_{24}$-Skyrme model, where only the kinetic $\mathcal{L}_2$ and Skyrme term $\mathcal{L}_4$ are considered.
Applying the scaling (Derrick) transformation $L\to \lambda L$, which for this version of the Skyrme model is an exact transformation relating solutions at different densities (volume), we can prove that $K_0/E_0=1$, see Appendix \ref{derrick}.
For the full Skyrme model such a transformation leads to an approximate relation between the energies of solutions at different densities.
However, for not too large $M_\pi$ and $\lambda$ (which are the coupling constants of the potential $\mathcal{L}_0$ and the sextic therm  $\mathcal{L}_6$), the approximation is quite accurate, especially at the saturation point.
Thus, for these versions of the model the compression modulus must be even higher than that in the minimal Skyrme model.

When $\lambda$ grows significantly the scaling transformation becomes useless.
This is because of the fact that in the the BPS Skyrme model, i.e., where only the potential and the sextic term are kept
\begin{equation}
    \mathcal{L}_{\textup{BPS}}=\mathcal{L}_6+\mathcal{L}_0
\end{equation}
the Skyrme field is a perfect fluid matter \cite{Naya_2014}.
Importantly, applying external pressure (squeezing skyrmionic matter) is not equivalent to the uniform rescaling of the solution.
On the contrary, regions with smaller densities are more squeezed than regions with higher density.
As a result, the compression modulus is zero for any potential which provides a nonzero mass of pions \cite{Naya_2014}.
Therefore, a near-BPS model, where the usual quadratic and quartic terms are added, may be a solution for the compression modulus problem.
However, the values of the parameters to reach the near-BPS regime must be extremely large, rendering the numerical computations difficult and the physical interpretation cumbersome.
Anyway, this opens a possible path for resolution of the compression modulus problem.

In the subsequent part of the paper we will follow another line and show that inclusion of further degrees of freedom, here $\rho$-mesons, reduces the stiffness of the pionic Skyrme model.
The reason for this originates in an observation that the Yang-Mills theory in (4+0) dimensions is equivalent to the 3-dimensional massless Skyrme model with an infinite tower of vector mesons \cite{Sutcliffe:2010et}.
This identification is obviously kept on the level of solutions with self-dual instantons of charge $N$ being "decomposed" into Skyrmions with baryon charge $B=N$ assisted by the vector mesons \cite{Sutcliffe:2010et}. 

The first consequence of this equivalence is that skyrmions with infinitely many mesons have zero binding energy as the self-dual instantons saturate the pertinent linear topological bound on the energy.
Following that small binding energies arise in a truncated model where only finite number of mesons are kept \cite{Sutcliffe:2011ig}.
This provides a very natural path of resolution of the too large binding energy problem in the Skyrme model. 

However, the instanton-skyrmion equivalence has another very important consequence.
Since the Yang-Mills theory in four dimensions is conformally invariant, the energy of the self-dual instanton does not depend on the scale.
Thus, it costs zero energy to change the size of the instantons.
This finally results in zero compression modulus.
Again, a non-zero, but physically small, value of the compression modulus may be found if a truncated skyrmion-meson theory is considered. 

It is interesting to notice that, one of the $\rho$-pion interaction terms, which we in fact are going to keep in our work, effectively reduces the quartic Skyrme term.
In the limit of maximal coupling it even kills the Skyrme term completely. 

Finally, we remark that the $\omega$-Skyrme model considered in \cite{Leask_2023} seems to be conceptually closer to a type Skyrme model with $\mathcal{L}_6$ included than to the $\rho$-meson extension.
This is because of the fact that the $\omega$-meson coupling effectively introduces a topological $B_\mu^2$ term. 


\section{The $\rho$-Skyrme model}

In general, we are interested in topological solitons in a $(3+1)$-dimensional Minkowski spacetime $\Sigma = \mathbb{R} \times \mathbb{R}^3$ with constant coefficient metric $\eta$ and metric signature $(-+++)$.
Then, the Skyrme field is a single scalar field $\varphi: \mathbb{R} \times \mathbb{R}^3 \rightarrow \SU(2)$, where the pion fields $\vec{\pi}=(\pi_1,\pi_2,\pi_3)$ are disguised into the Skyrme field $\varphi$ via
\begin{equation}
    \varphi=\begin{pmatrix}
        \sigma + i\pi_3 & i\pi_1 + \pi_2 \\
        i\pi_1 - \pi_2 & \sigma - i\pi_3 
    \end{pmatrix} \in \SU(2),
\end{equation}
with the $\sigma$ field imposing the constraint $\sigma + \pi_i\pi_i = 1$.
We will consider the $\rho$-meson in its standard version, i.e. being a massive non-Abelian field.
In the literature there is another {\it inequivalent} way to introduce the $\rho$-meson within the pionic effective Lagrangian.
Namely, via a constrained $2\times 2$ four vector \cite{Adkins_1986}.
We choose a non-constrained, Lie algebra valued formulation since it is computationally (and in particular numerically) simpler as it does not contain non-physical degrees of freedom.
We can thus express the $\rho$-meson as an $\su(2)$-valued vector $R_\mu$, where $R_\mu=i \rho^a_\mu \tau^a$ with $\rho^a_\mu \in \mathbb{R}$ and $\tau^a$ are the Pauli spin matrices.

The model we are interested in is the massive Skyrme model coupled to the $\rho$-meson, described by the Lagrangian
\begin{equation}
\begin{split}
    \mathcal{L} = - \frac{1}{8\hbar^3}F_\pi^2 m_\pi^2 \Tr\left( \textup{Id} - \varphi \right) + \frac{F_\pi^2}{16\hbar}\eta^{\mu\nu} \Tr(L_\mu L_\nu) + \frac{\hbar}{32e^2}\eta^{\mu\alpha}\eta^{\nu\beta} \Tr\left( [L_\mu, L_\nu] [L_\alpha, L_\beta] \right) \\ -\frac{m_\rho^2}{4\hbar^3} \eta^{\mu\nu} \Tr\left(R_\mu^\dagger R_\nu\right) - \frac{1}{8\hbar} \eta^{\mu\alpha}\eta^{\nu\beta}\Tr\left(R_{\mu\nu}^\dagger R_{\alpha\beta}\right) + \frac{1}{2} \eta^{\mu\beta}\eta^{\nu\gamma} \alpha\Tr\left( R_{\mu\nu}[L_\beta, L_\gamma] \right)
\end{split},
\label{eq: Main Lagrangian}
\end{equation}
where $L_\mu= \varphi^{-1}\partial_\mu \varphi$ is an $\su(2)$-valued left current and $R_{\mu\nu} = \partial_\mu R_\nu - \partial_\nu R_\mu$. 
The parameters of the model are the pion decay constant $F_\pi=129\,\textup{MeV}$, the pion mass $m_\pi=138\,\textup{MeV}$, the $\rho$-meson mass $m_\rho=775\,\textup{MeV}$, the dimensionless Skyrme parameter $e$, and the $\rho\pi\pi$ coupling constant $\alpha$.
Also, $\hbar = 197.33\,\textup{MeV fm}$ is the reduced Planck constant.

The above formulation is very well physically motivated, that is, it emerges via a holographic like construction from a dimensional reduction of the four dimensional static Yang-Mills theory with $\SU(2)$ gauge group \cite{Sutcliffe:2010et}. 
Importantly, this derivation dictates also interaction terms and their coupling constants.
In principle, there are six different interaction terms.
In the current paper, due to the considerable complexity of the resulting model, we consider a simplified version of this construction and take into account only one of them.
We also assume that its coupling constant is a free parameter rather than an instanton related fixed constant.
This is a consistent low energy truncation as we leave the terms which are the lowest (first) order in the $\rho$-meson field.
Furthermore, the $\rho\pi\pi$ vertex
\begin{equation}
    \mathcal{L}_{\rho\pi\pi} = 2\alpha m_\rho^2 \epsilon_{abc} \rho^{\nu}_c \pi_a \partial_{\nu} \pi_b
\end{equation}
is included in this truncation.

For our numerical computations we will use a rescaled theory, with the energy scale $\tilde{E}=F_\pi/(4e)$ (MeV) and the length scale $\tilde{L}=2\hbar / (eF_\pi)$ (fm).
We also further rescale the $\rho$-meson field by $R_\mu \to F_\pi R_\mu$.
Then the rescaled $\rho$-Skyrme Lagrangian in dimensionless Skyrme units is given by
\begin{equation}
\begin{split}
    \mathcal{L} = -M_\pi^2 \Tr\left( \textup{Id} - \varphi \right) + \frac{1}{2}\eta^{\mu\nu} \Tr(L_\mu L_\nu) + \frac{1}{16}\eta^{\mu\alpha}\eta^{\nu\beta} \Tr\left( [L_\mu, L_\nu] [L_\alpha, L_\beta] \right) \\ + 2M_\rho^2 \eta^{\mu\nu} \Tr\left(R_\mu R_\nu\right) + \eta^{\mu\alpha}\eta^{\nu\beta}\Tr\left(R_{\mu\nu}R_{\alpha\beta}\right)  + 2 \alpha e \eta^{\mu\alpha}\eta^{\nu\beta} \Tr\left( R_{\mu\nu} [L_\alpha, L_\beta] \right)
    \end{split},
\end{equation}
where the rescaled pion and $\rho$-meson masses are, respectively,
\begin{equation}
    M_\pi = \frac{2 m_\pi}{e F_\pi}, \quad M_\rho = \frac{2m_\rho}{e F_\pi}.
\end{equation}
The associated energy-momentum tensor (in dimensionless Skyrme units) can be obtained via the Hilbert prescription, which yields
\begin{equation}
    \begin{split}
        T_{\mu\nu} = -\Tr(L_\mu L_\nu) - 4 M_\rho^2 \Tr\left(R_\mu R_\nu\right) - \frac{1}{4} \eta^{\alpha\beta}\Tr([L_\mu,L_\alpha][L_\nu,L_\beta]) - 4 \eta^{\alpha\beta}\Tr\left(R_{\mu\alpha}R_{\nu\beta}\right) \\
        - 8\alpha e \eta^{\alpha\beta}\Tr\left( R_{\mu\alpha} [L_\nu,L_\beta]\right) + \eta_{\mu\nu} \mathcal{L}
    \end{split}.
\end{equation}
From the time-like part of the energy-momentum tensor we obtain the static energy functional of theory, that is
\begin{equation}
    \begin{split}
        E = \int \textup{d}^3x \, \sqrt{-\eta} \left\{ M_\pi^2 \Tr\left( \textup{Id} - \varphi \right) -\frac{1}{2}\eta^{ij} \Tr(L_i L_j) - \frac{1}{16}\eta^{ia}\eta^{jb} \Tr\left( [L_i, L_j] [L_a, L_b] \right) \right. \\ \left. -2M_\rho^2\eta^{ij}\Tr\left(R_i R_j\right) - \eta^{ia}\eta^{jb}\Tr\left(R_{ij}R_{ab}\right) -2\alpha e \eta^{ia}\eta^{jb}\Tr\left(R_{ij}[L_a,L_b] \right) \right\}
    \end{split}.
    \label{eq: Main energy}
\end{equation}

The first term in the energy \eqref{eq: Main energy} is the standard pion mass potential, the second is the kinetic term and the third is the Skyrme term, which corresponds to the four pion interaction.
These three terms makeup the standard massive $\mathcal{L}_{024}$-Skyrme model.
There exists a generalization of the massive $\mathcal{L}_{024}$-model that includes the addition of the sixth order sextic term \cite{Jackson_1985}
\begin{equation}
    \mathcal{L}_6 = -\pi^4 \lambda^2 \eta^{\mu\nu} \mathcal{B}_\mu \mathcal{B}_\nu,
\end{equation}
which yields an $\omega$-meson like repulsion on short distances, while also allowing the fourth order Skyrme term to describe scalar meson effects.
The constant $\lambda$ is related to the $\omega$-meson mass $m_\omega$ and the coupling constant $\beta_\omega$ of the $\omega$-meson via $\lambda^2 = \beta_\omega^2 \hbar^3/(2\pi^4 m_\omega^2)$ \cite{Wereszczynski_2015}.

Inclusion of the sextic term, in general, stiffens the nuclear matter equation of state, especially at higher densities, and increases the maximal obtainable masses of neutron stars within the theory to be more in line with observed data \cite{Adam_2020}.
However, the caveat is that it also may increase the compression modulus. Furthermore, this term is numerically quite challenging. 
For this reason, we will only be concerned with the massive $\mathcal{L}_{024}$-model as we are interested in the effects of the $\rho$-meson on the Skyrme model and, in particular, on the compression modulus (i.e., compressibility at the saturation point where the role of sextic is expected to be subleading).

Before discussing skyrmion solutions, let us observe that the model possesses a topological energy bound.
This follows from the observation that a part of the $\rho$-meson Lagrangian can be rewritten as 
\begin{equation}
    \begin{split}
        \int_{\mathbb{R}^3} \textup{d}^3 x \left( \Tr |R_{ij}|^2 - 2e\alpha 
    \Tr \left( R_{ij} [L_i,L_j]\right) \right) 
    =\int_{\mathbb{R}^3} \textup{d}^3 x \Tr \left( R_{ij} + e\alpha [L_i,L_j] \right) \left(R_{ij} +e\alpha [L_i,L_j] \right)^\dagger \\ + e^2\alpha^2 \int_{\mathbb{R}^3} \textup{d}^3 x \Tr  [L_i,L_j]^2.
    \end{split} 
\end{equation}
where for simplicity flat base space has been assumed.
The last term is the quartic Skyrme term with negative sign.
Hence, the full static energy reads
\begin{align}
    E\, &=\int_{\mathbb{R}^3} \textup{d}^3x \left( \frac{1}{2} \Tr |L_i|^2 + \frac{\tilde{c}_4}{4} \Tr |[L_i,L_j]|^2 + M_\pi^2 \Tr (\mbox{Id}-\phi) \right) \nonumber \\
    \,& + \int_{\mathbb{R}^3} \textup{d}^3 x \left(2M_\rho^2 \Tr |R_i|^2 + \Tr \left| R_{ij} +\frac{e\alpha}{2} 
    [L_i,L_j] \right|^2 \right) \nonumber \\
    \, & \geq \int_{\mathbb{R}^3} \textup{d}^3 x  \left( \frac{1}{2} \Tr |L_i|^2 + \frac{\tilde{c}_4}{4} \Tr |[L_i,L_j]|^2 \right) \geq 24 \pi^2 \sqrt{\tilde{c}_4} |B|,
\end{align}
where in the last step we used the usual Faddeev--Skyrme bound.
This computation makes sense only if the new coupling constant in front of quartic term is positive $\tilde{c}_4 = \frac{1}{4}-4e^2\alpha^2 >0$.
Thus we arrive at an upper bound for the coupling constant $\alpha$ 
\begin{equation}
    c_\alpha := \alpha e \leq \frac{1}{4}.
\end{equation}
Larger values would destabilize skyrmions as they lead to non-positive definite energy integral.
This analysis shows also the role of $\rho$-mesons.
They effectively reduce the coupling constant of the quartic term and, in an extremal case, can make it arbitrary small. 

Now we turn our attention to topological solitons in the $\rho$-Skyrme model, that is, skyrmions.
These are solutions of the full non-linear Euler--Lagrange equations associated to the static energy functional \eqref{eq: Main energy}.
If we consider the model on a $3$-torus, and also allow the fundamental period lattice of the torus to vary, we obtain crystalline configurations of skyrmions -- skyrmion crystals.
This is essential for understanding phases and phase transitions of nuclear matter within the Skyrme model.
However, before we turn our attention to skyrmion crystals, we first detail the $B=1$ hedgehog skyrmion within the $\rho$-Skyrme model, which describes the nucleon, and also some light nuclei.


\subsection{The $B=1$ skyrmion: hedgehog ansatz}
\label{subsec: Hedgehog ansatz}

The $B=1$ skyrmion is necessarily spherically symmetric and the pion fields $\vec{\pi}$ point radially outwards, earning the moniker the ``hedgehog'' skyrmion.
This can be obtained by use of a rational map ansatz (RMA) \cite{Houghton_1998}.
The RMA is carried out in a local spherical coordinate system  $(r,\theta,\phi)$, and the Skyrme field ansatz is defined by \cite{Houghton_1998}
\begin{align}
    \varphi(r,\theta,\phi) = \left( \cos f(r), \sin f(r) \vec{n}_W(\theta,\phi) \right),
\end{align}
where the unit vector $\vec{n}_W$ in terms of the rational map $W:\mathbb{C} \rightarrow \mathbb{C}$ is given explicitly by
\begin{equation}
    \vec{n}_W = \left( \frac{2\Re(W)}{1+|W|^2}, \frac{2\Im(W)}{1+|W|^2}, \frac{1-|W|^2}{1+|W|^2} \right).
\end{equation}
Here $f:[0,\infty)\rightarrow\mathbb{R}$ is a radial profile function and we identify the rational map domain $S^2 \cong \mathbb{C} \cup \{\infty\}$ via the Riemman sphere coordinate $z=\tan(\theta/2)\exp(i\phi)$.
For suitable choices of profile functions and rational maps, the ansatz produces excellent approximations to the true skyrmions, with the correct symmetries as well.
In terms of the RMA, the hedgehog ansatz is defined by the rational map $W(z)=z$, which gives the unit normal
\begin{equation}
    \vec{n}_{H} = \left( \sin\theta \cos\phi, \sin\theta \sin\phi, \cos\theta \right).
\end{equation}
The corresponding standard hedgehog Skyrme field is given by
\begin{equation}
    \varphi_H = \begin{bmatrix}
        \cos f(r)+i \cos\theta \sin f(r) & \sin\theta \sin\phi \sin f(r)+i \sin\theta \cos\phi \sin f(r) \\
        -\sin\theta \sin\phi \sin f(r)+i \sin\theta \cos\phi \sin f(r) & \cos f(r)-i \cos\theta \sin f(r) \\
    \end{bmatrix}.
\label{eq: Hedgehog - Skyrme}
\end{equation}
As a result of the hedgehog symmetry and the intrinsic parity of the $\rho$-mesons, the most general form for the spin-1 $\rho$-field is \cite{Meissner_1986}
\begin{equation}
    R^{H}_i = i \rho^a_i \tau^a, \quad \rho^a_i = \epsilon_{aij} n_{H}^j \xi(r).
\label{eq: Hedgehog - Rho}
\end{equation}

The unknown profile functions $f(r)$ and $\xi(r)$ are obtained by substituting the ansatz \eqref{eq: Hedgehog - Skyrme} and \eqref{eq: Hedgehog - Rho} into the static energy~\eqref{eq: Main energy}, and minimizing the static nucleon energy
\begin{equation}
    E_1 = 4\pi\int_{0}^{\infty} \textup{d}r \, \mathcal{E}_H(r),
\end{equation}
where the hedgehog energy density is
\begin{align}
    \mathcal{E}_{H}(r) = \, & 2 M_{\pi }^2 r^2 \left[1-\cos f(r)\right] + r^ 2f'(r)^2 + 2 \sin^2f(r) + \sin ^2f(r)  \left[2 f'(r)^2+\frac{\sin^2f(r)}{r^2}\right] \nonumber \\
    \, &  + 8 r^2 M_{\rho }^2 \xi (r)^2 + 8\left[3 \xi (r)^2 + 2 r\xi (r) \xi '(r) + r^2 \xi '(r)^2\right] \nonumber \\
    \, & + \frac{1}{r} 32 \alpha e \sin f(r) \left[r f'(r) \cos f(r) \left\{r \xi'(r) + \xi (r)\right\} + \xi (r) \sin f(r)\right].
\end{align}
That is, the profile functions are determined by solving the coupled pair of ODEs,
\begin{equation}
    \begin{split}
        2 M_{\pi }^2 r^2 \sin f(r) + 2\sin2 f(r) - 2r^2 f''(r)
        - 4\sin f(r)\left\{f''(r) \sin f(r)+f'(r)^2 \cos f(r)\right\}
        \\ 
        - 4 r f'(r) + \frac{4\sin^3f(r) \cos f(r)}{r^2} + 16 \alpha e \sin2 f(r)  \left[ \frac{2 \xi(r)}{r} -  \left\{ r \xi''(r)+2 \xi '(r)\right\}\right] = 0
    \end{split}
    \label{eq: f EoM}
\end{equation}
and
\begin{equation}
    \begin{split}
        16 \alpha e r \left[- f''(r) \sin2 f(r) - 2 f'(r)^2 \cos2 f(r) \right. \left. + \frac{2 \sin^2f(r)}{r^2}\right] \\+ 16 \left[ \xi(r) \left\{ r^2 M_{\rho }^2 + 2 \right\} - r^2 \xi''(r) - 2 r \xi'(r)\right] = 0
    \end{split}.
    \label{eq: xi EoM}
\end{equation}

\begin{figure}[h]
    \centering
    \begin{subfigure}[b]{0.4\textwidth}
        \includegraphics[width=\textwidth]{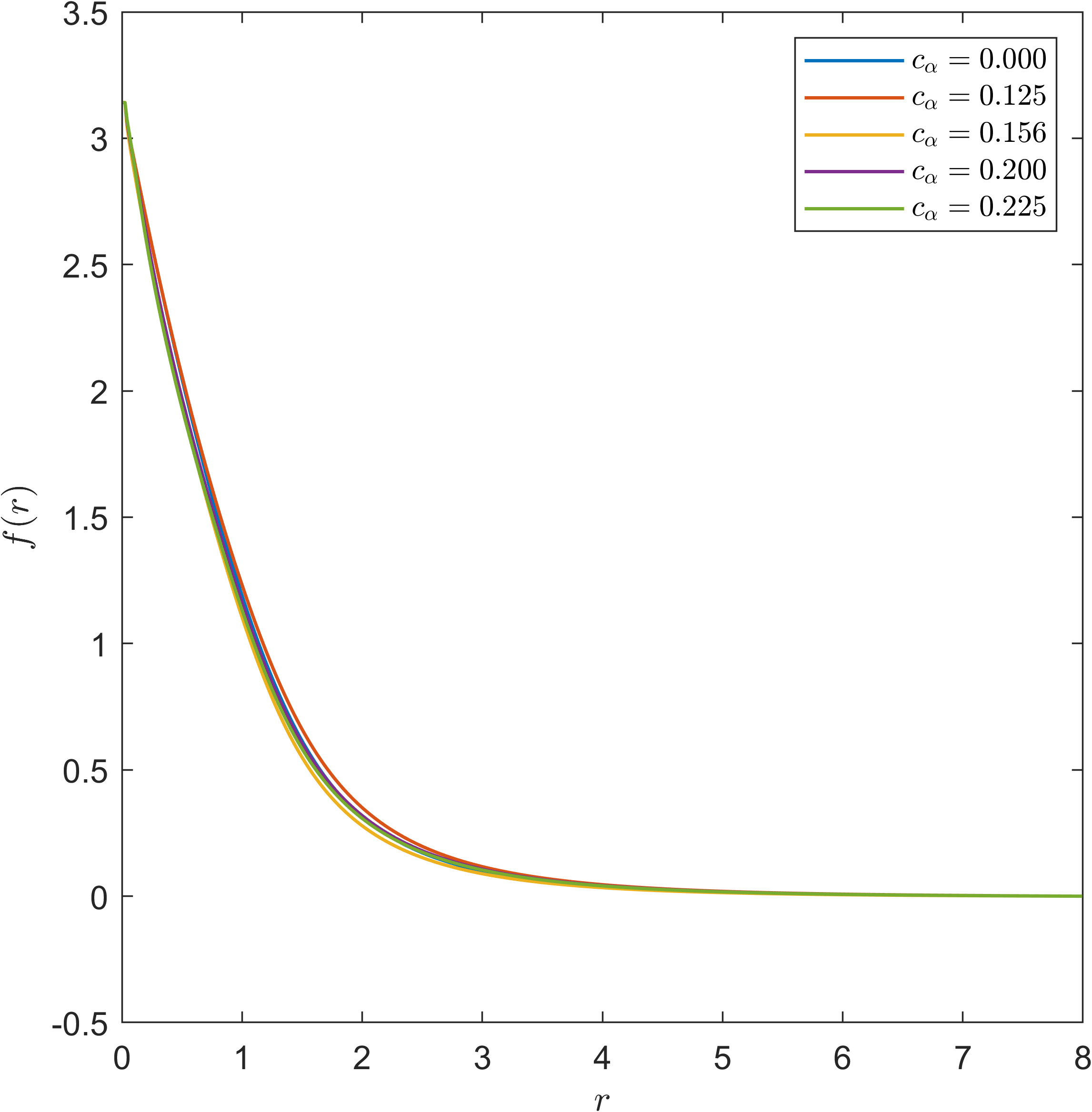}
        \caption{}
        \label{fig: Baby Skyrmions on R2 - EP Charge-1 (b)}
    \end{subfigure}
    ~
    \begin{subfigure}[b]{0.4\textwidth}
        \includegraphics[width=\textwidth]{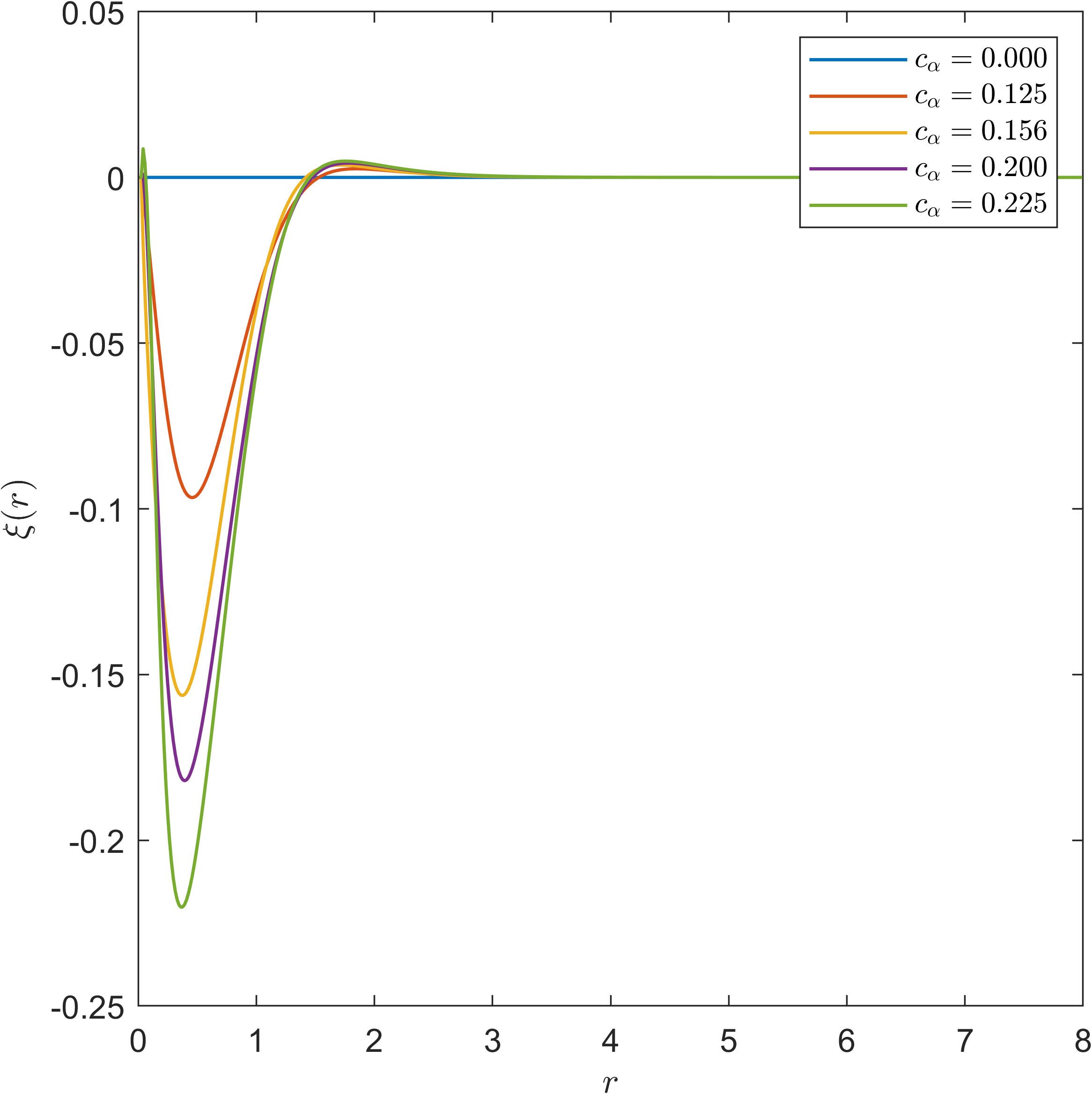}
        \caption{}
        \label{fig: Baby Skyrmions on R2 - EP Charge-1 (b)}
    \end{subfigure}
    \\
    \begin{subfigure}[b]{0.4\textwidth}
        \includegraphics[width=\textwidth]{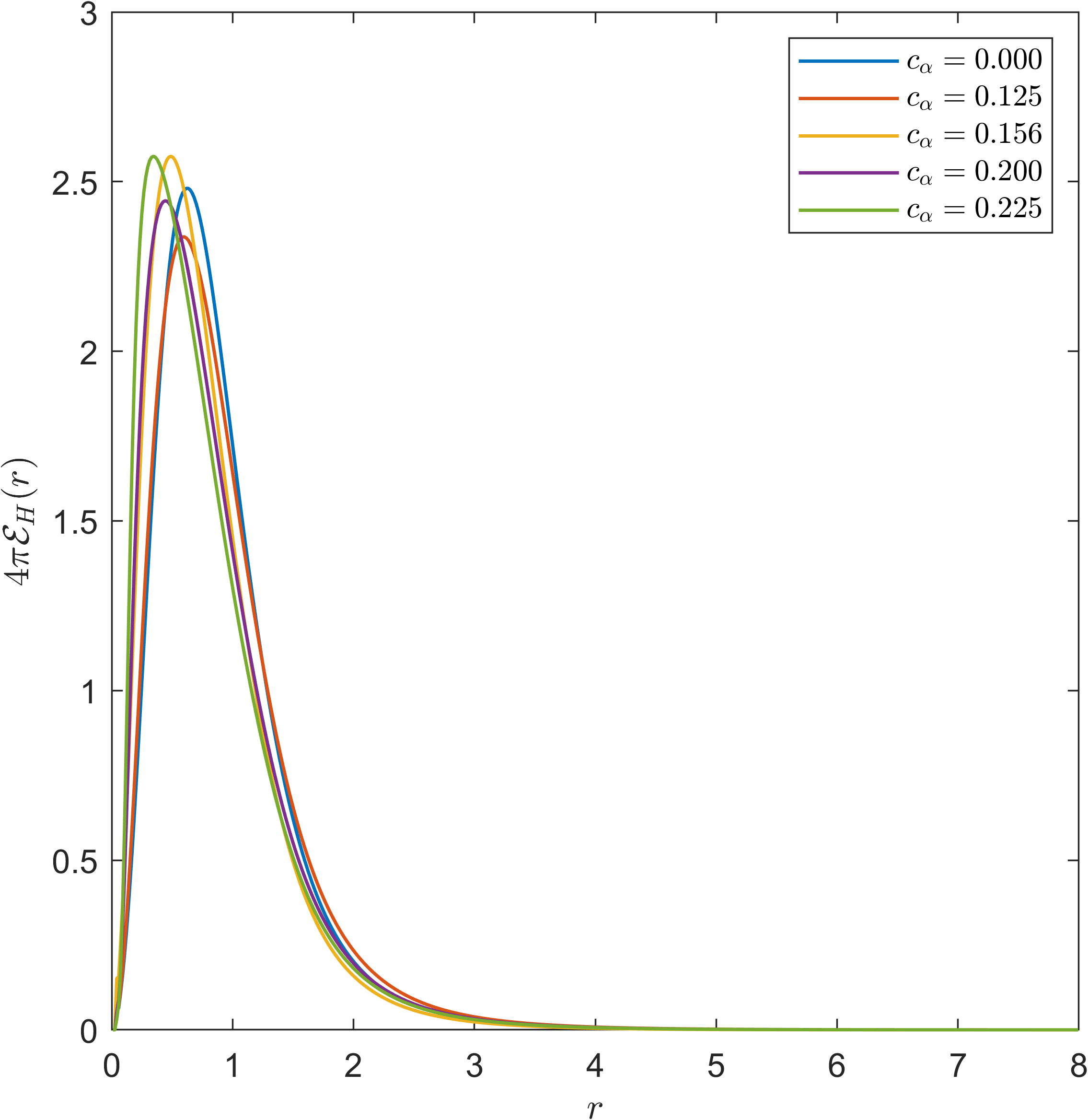}
        \caption{}
        \label{fig: Baby Skyrmions on R2 - EA Charge-1 (a)}
    \end{subfigure}
    \caption{Plots for various coupling constants $c_\alpha$ of (a) the Skyrme radial profile functions $f(r)$, (b) the $\rho$-meson radial profile $\xi(r)$ and (c) the energy density profile $\mathcal{E}_{H}(r)$ obtained by minimizing the static nucleon mass $E_1$ via a multiple shooting method. The chiral angle $f(r)$ is related to the pion field and $\xi(r)$ is related to the spatial components of the $\rho$-meson.}
    \label{fig: Profile functions}
\end{figure}

To ensure the soliton is bounded in space, the appropriate boundary conditions are \cite{Adkins_1986}
\begin{equation}
    f(0)=\pi, \quad f(\infty)=0, \quad \xi(0)=0, \quad \xi(\infty)=0.
\label{eq: Profile BCs}
\end{equation}
The resulting profile functions $f(r)$ and $\xi(r)$ are obtained by applying a multiple shooting method to the coupled system \eqref{eq: f EoM}-\eqref{eq: xi EoM} with the fixed boundary conditions \eqref{eq: Profile BCs}.
For increasing values of the coupling constant $\alpha$, the energy minimizing profile functions are shown in Fig.~\ref{fig: Profile functions} alongside the energy density.
We remark that the profile function of the Skyrme field is very weakly affected by the $\rho$-meson, enjoying a sort of universality as recently observed in \cite{Manton2024}. 

\newpage

\subsection{Multi-skyrmions with $\rho$-mesons}
\label{subsec: Multi-skyrmions with rho mesons}

For higher charge skyrmions $B>1$, there is no simplification to a one dimensional problem as there is for the hedgehog solution.
We must solve the full non-linear Euler--Lagrange field equations.
To solve the field equations we use arrested Newton flow, an accelerated gradient descent method with flow arresting, with the RMA as an initial configuration.
The well-known rational maps of high symmetry which we use as initial configurations for the numerical relaxation are detailed in \cite{Houghton_1998}.

\begin{figure}[t]
    \centering
    \includegraphics[width=0.6\columnwidth]{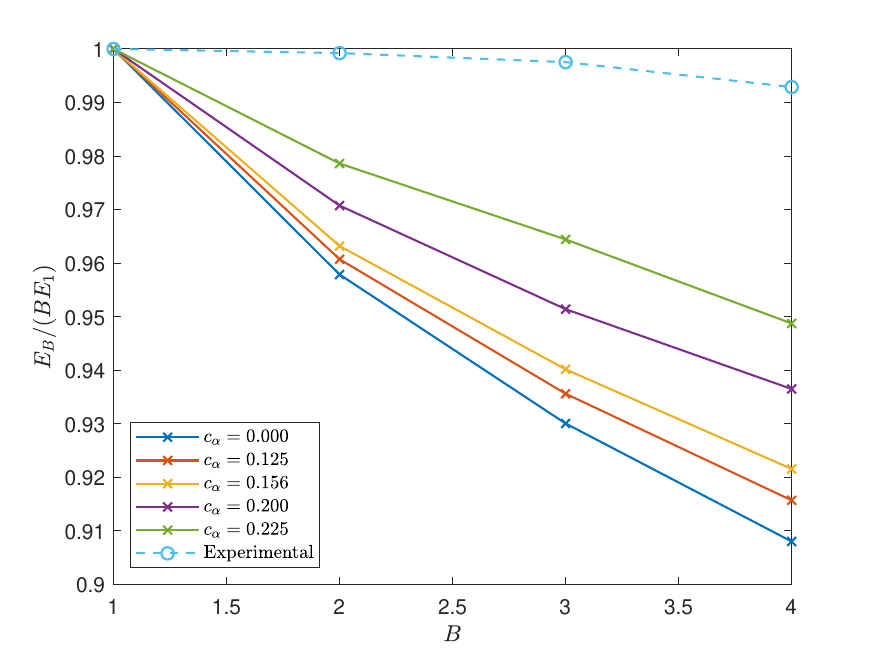}
    \caption{Comparison of the normalized energies per baryon of the multi-skyrmion solutions up to $B=4$. It can clearly be seen that the binding energies decrease as the coupling is increased.} 
    \label{fig: Binding energies}
\end{figure}

In general, the initial configuration is not a minimizer and so it swaps its potential energy for kinetic energy as it evolves.
During the evolution we check to see if the energy is increasing.
If the energy is indeed increasing, we take out all the kinetic energy in the system by setting the velocity to zero and restart the flow (this is the arresting criteria).
Naturally the field will relax to a local, or global, minimum in some potential well.
The evolution then terminates when every component of the energy gradient is zero within some specified tolerance, $\textup{tol}=10^{-5}$.

Explicitly, we are solving Newton's equations of motion for a particle on the discretised configuration space $\mathcal{C}$ with potential energy $E_{\textup{dis}}$.
That is, we are solving the coupled system of 1st order ODEs
\begin{equation}
    \dot{\psi}^\mu = -\frac{\delta \mathcal{E}}{\delta \varphi^\mu}, \quad \psi^\mu = \dot{\varphi}^\mu,
\label{eq: Numerical minimisation procedure - 2nd order ODEs}
\end{equation}
with initial configuration $\varphi(0)=\varphi_{\textup{RMA}}$ and initial velocity $\psi(0)=\dot{\varphi}(0)=0$.
We implement a fourth order Runge--Kutta method to solve this coupled system.
In conjunction, we do the exact same for the $\rho$-meson fields:
\begin{equation}
    \dot{P}^a_b = -\frac{\delta \mathcal{E}}{\delta \rho^a_b}, \quad P^a_b = \dot{\rho}^a_b,
\end{equation}
with initial configuration $\rho^a_b=0$ and initial velocity $P^a_b(0)=\dot{\rho}^a_b(0)=0$.
We also implement a fourth order Runge--Kutta method to solve this coupled system.
The arresting criteria is implemented as follows.
If $E(t + \delta t) > E(t)$, we take out all the kinetic energy in the Skyrme system by setting $\psi(t + \delta t)=0$ and $P(t + \delta t)=0$, and restart the flow for the Skyrme field and the $\rho$-mesons.

\begin{figure}[t]
    \centering
    \begin{subfigure}[b]{0.2\textwidth}
        \includegraphics[width=\textwidth]{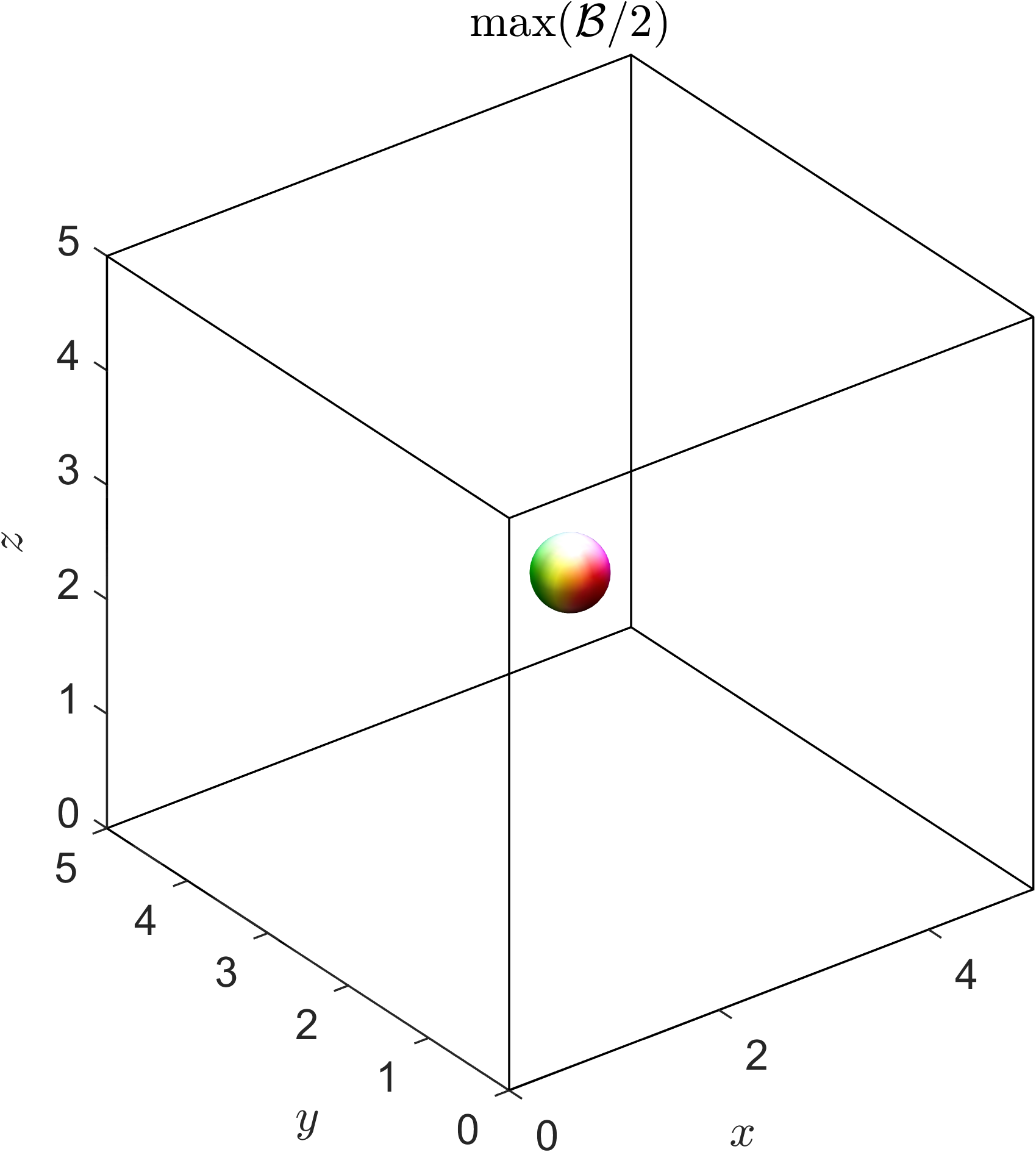}
        \caption{$B=1$}
        \label{fig: B=1}
    \end{subfigure}
    ~
    \begin{subfigure}[b]{0.2\textwidth}
        \includegraphics[width=\textwidth]{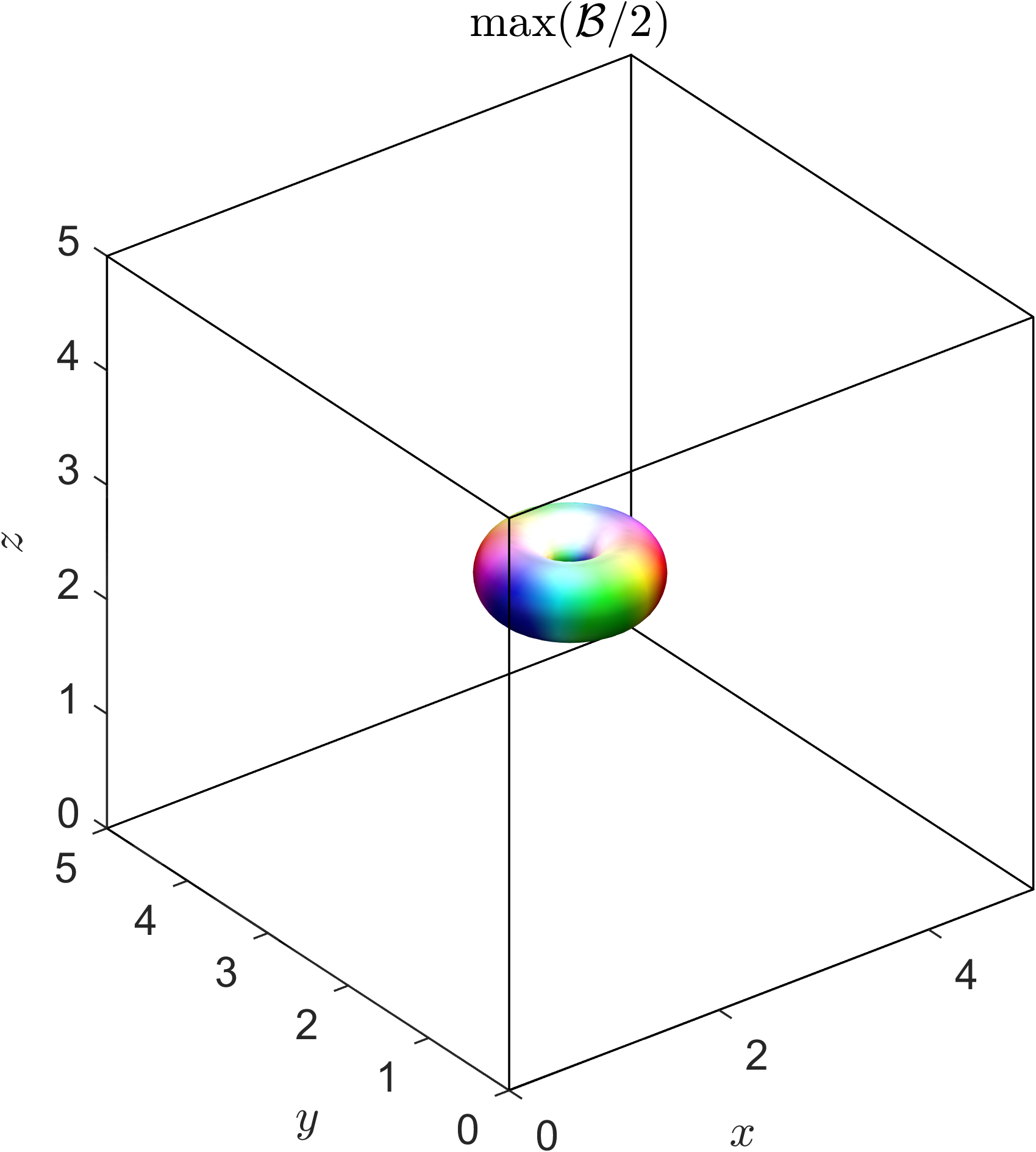}
        \caption{$B=2$}
        \label{fig: B=2}
    \end{subfigure}
    ~
    \begin{subfigure}[b]{0.2\textwidth}
        \includegraphics[width=\textwidth]{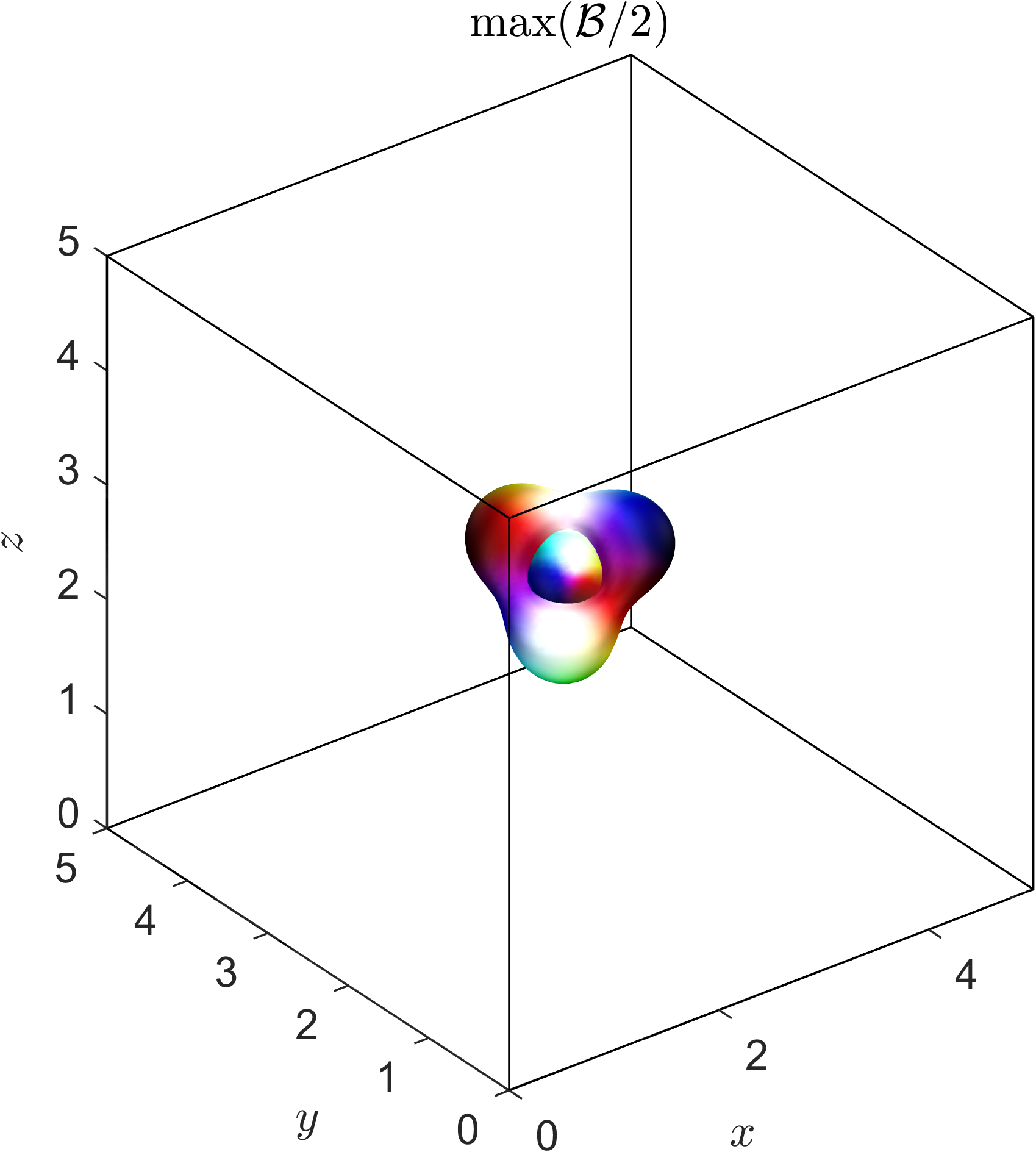}
        \caption{$B=3$}
        \label{fig: B=3}
    \end{subfigure}
    ~
    \begin{subfigure}[b]{0.2\textwidth}
        \includegraphics[width=\textwidth]{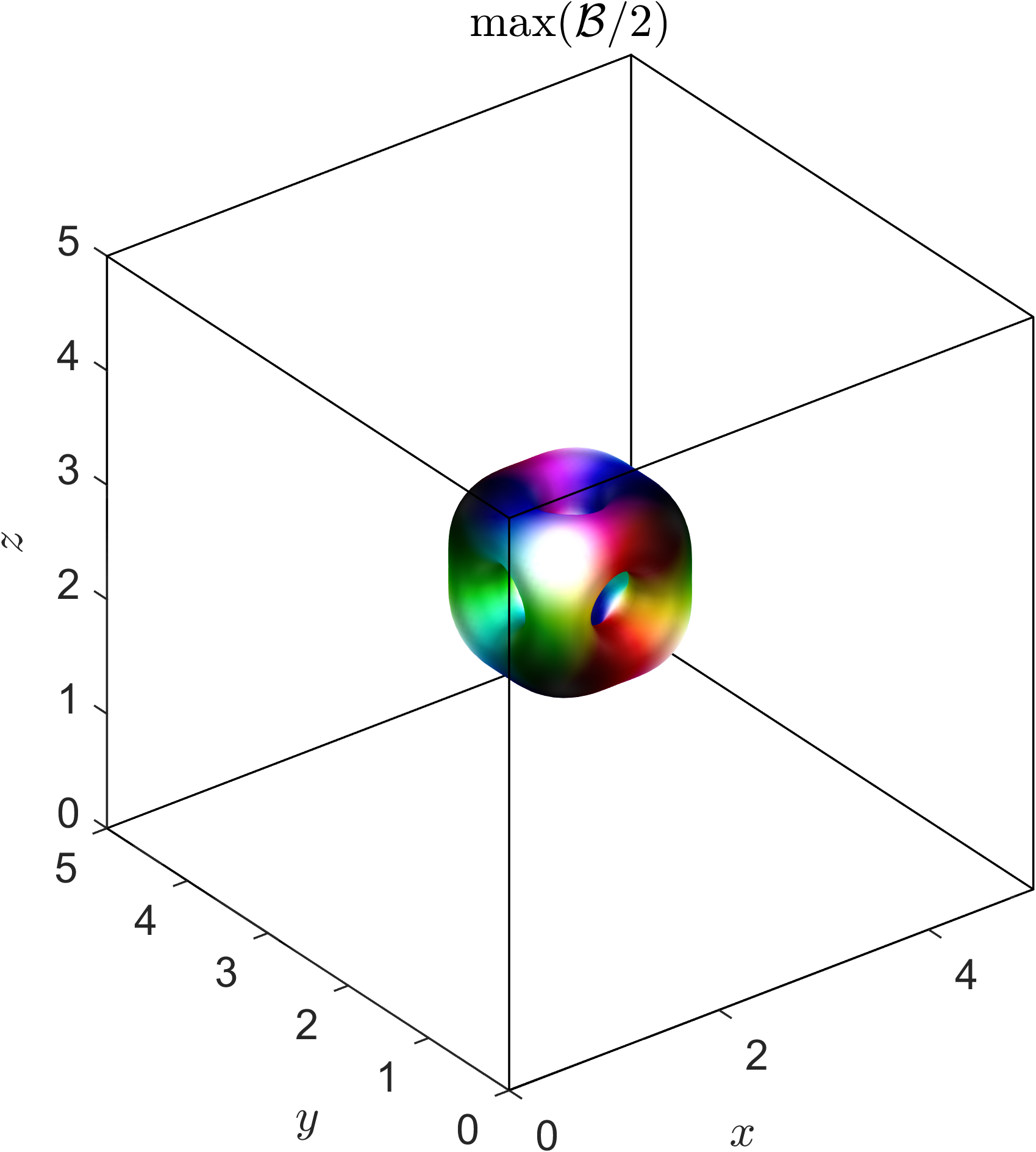}
        \caption{$B=4$}
        \label{fig: B=4}
    \end{subfigure}
    \caption{Level set plots of the baryon density of multi-skyrmions coupled to $\rho$-mesons for $B=1,2,3,4$ for the choice of coupling $c_\alpha=0.200$.}
    \label{fig: Baryon densities}
\end{figure}

The resulting binding energies as a function the coupling constant $c_\alpha$ are plotted in Fig.~\ref{fig: Binding energies}.
Alongside this, baryon density plots for representative $c_\alpha=0.200$ are shown in Fig.~\ref{fig: Baryon densities}.
It is apparent that the binding energies decrease as the coupling $c_\alpha$ is increased.

\section{Nuclear matter in the $\rho$-Skyrme model}
\label{sec: Light Nuclei}

The application of the skyrmion crystals to the analysis of nuclear matter was first studied by Klebanov \cite{Klebanov_1985} in the context of the massless $\mathcal{L}_{24}$-model.
This was carried out for the simple cubic crystal of hedgehog skyrmions ($\textup{SC}_1$) with unit cell charge $B_{\textup{cell}}=8$, where it was observed that this crystal undergoes a phase transition to a body centered cubic crystal of half skyrmions ($\textup{BCC}_{1/2}$) at higher densities \cite{Goldhaber_1987}.
It was later found that a crystal with lower energy than the $\textup{BCC}_{1/2}$ exists, a simple cubic crystal of half skyrmions ($\textup{SC}_{1/2}$) with unit cell charge $B_{\textup{cell}}=4$ \cite{Kugler_1988,Castillejo_1989}.
Baskerville took Klebanov's method further and applied it to the $\textup{SC}_{1/2}$ crystal, and the state corresponding to a neutron crystal was identified \cite{Baskerville_1996}.

The phase structure of the $\mathcal{L}_{24}$-Skyrme model has been studied by Jackson and Verbaarschot \cite{Jackson_1988}.
This has been further investigated in the generalized $\mathcal{L}_{0246}$-Skyrme model \cite{Shnir_2017}. 
Two candidates have been previously proposed as the minimal $E/B$ crystal for $\mathcal{L}_{024}$-skyrmions, these are the $\textup{SC}_{1/2}$ crystal and the $\alpha$-particle crystal \cite{Feist_2012}.
Many skyrmions can be constructed as chunks of the infinite crystal \cite{Baskerville_1996_2,Feist_2013} and many have been built from $\alpha$-particles \cite{Battye_2007}.
The phase transition between the $\alpha$-particle and the $\alpha$-particle lattice has been investigated in both the $\mathcal{L}_{24}$-model \cite{SilvaLobo_2010} and $\mathcal{L}_{0246}$-model \cite{Adam_2022}.
This $B=4$, or $\alpha$-particle, cluster picture supports the $\alpha$-particle model of nuclei in which medium to large skyrmions are composed of $B=4$ skyrmions in many arrangements \cite{Feist_2013,Battye_2007}.
This $\alpha$-clustering model has been pretty successful in the predicting the energy spectrum of states of Carbon-12 \cite{Lau_2014,Rawlinson_2018} and Oxygen-16 \cite{Halcrow_2017,Manton_2019}.
For a recent review of the application of skyrmion crystals to the study of dense nuclear matter see, e.g., \cite{Huidobro_2023}.

Recently, the crystalline structure of skyrmion matter was studied more rigorously in the massive $\mathcal{L}_{024}$-model \cite{Leask_2023} and the generalized $\mathcal{L}_{0246}$-model \cite{Leask_2024}.
In both studies, four crystal solutions were found with unit cell charge $B_{\textup{cell}}=4$.
Two of these solutions were known before and have cubic lattice geometry, these are the $\textup{SC}_{1/2}$ and $\alpha$-particle crystals.
The two new solutions have non-cubic period lattices and take the form of a chain and a multiwall configuration.
It was observed that the ground state crystalline configuration is the multiwall solution in the Skyrme model without vector mesons.
This was found to be true at all densities and for various choices of the coupling constants.

However, when the theory is coupled to the $\omega$-meson \cite{Leask_Harland_2024}, the ground state crystal configuration is dependent upon the underlying parameters of the theory.
A switching in the energy ordering of the four crystal solutions was observed, with the multiwall crystal not always being the ground state configuration.
Before we can compute the compression modulus, we first need to determine the ground state crystal in this theory.


\subsection{Varying the metric}
\label{subsec: Varying the metric}

In order to be able to address the compression modulus problem, we first need to understand the crystalline nature of dense nuclear matter within the $\rho$-Skyrme framework.
To do this we follow the general methodology first proposed by Speight \cite{Speight_2014}.
That is, we study static Skyrme fields $\varphi:\mathbb{R}^3\rightarrow\SU(2)$ and $\rho$-meson fields $R_\mu:\mathbb{R}^3\rightarrow\su(2)$ that are periodic with respect to some $3$-dimensional period lattice 
\begin{equation}
    \Lambda = \left\{ n_1 \vec{X}_1 + n_2 \vec{X}_2 + n_3 \vec{X}_3: n_i \in \mathbb{Z} \right\}.
\end{equation}
We can equivalently interpret the domain of the fields as $\mathbb{R}^3/\Lambda$, and identify this with the unit $3$-torus $(\mathbb{T}^3,g)$ via the diffeomorphism
\begin{equation}
    F: \mathbb{T}^3 \rightarrow \mathbb{R}^3/\Lambda, \quad (x^1,x^2,x^3) \mapsto x^1\vec{X}_1 + x^2\vec{X}_2 + x^3\vec{X}_3.
\end{equation}
Then the resulting metric $g$ on $\mathbb{T}^3$ is the pullback of the metric $d$ by $F$, i.e.
\begin{equation}
    g = F^*d = g_{ij} \textup{d}x^i \textup{d}x^j, \quad g_{ij} = \vec{X}_i \cdot \vec{X}_j.
\end{equation}
Now, varying the period lattice $\Lambda_s$ with $\Lambda_0=\Lambda$ is equivalent to considering variations of the metric $g_s$ on $\mathbb{T}^3$ with $g_0=F^*d$.
Therefore, the energy minimized over variations $g_s$ of the domain metric is equivalent to determining the energy minimizing period lattice $\Lambda$.

Throughout, it will be convenient to use the non-linear $\sigma$ model (NL$\sigma$M) formulation of the model, where we identify $\SU(2) \cong S^3$ and write the Skyrme field as the unit $4$-vector $\varphi=(\sigma,\vec{\pi})$.
In this notation we identify $\varphi_0 \equiv\sigma$ and $\varphi_i\equiv\pi_i$.
So, now we can regard the Skyrme field as the map $\varphi: \mathbb{T}^3 \rightarrow S^3$ and the $\rho$-meson as the map $R_\mu: \mathbb{T}^3 \rightarrow \su(2)$.
Then, using the NL$\sigma$M formulation, the energy density can be written in index notation as
\begin{equation}
    \begin{split}
    \mathcal{E} =  2 M_\pi^2 (1-\varphi_0) + g^{ij}\partial_i\varphi_\mu \partial_j\varphi_\mu + \frac{1}{2}g^{ik}g^{jl}\left( \partial_i\varphi_\mu \partial_k\varphi_\mu \partial_j\varphi_\nu \partial_l\varphi_\nu - \partial_i\varphi_\mu \partial_l\varphi_\mu \partial_j\varphi_\nu \partial_k\varphi_\nu \right) \\ + 4 M_\rho^2g^{ij} \rho_i^a \rho_j^a + 2 g^{ik}g^{jl}\left( \partial_i \rho_j^a - \partial_j \rho_i^a \right) \left( \partial_k \rho_l^a - \partial_l \rho_k^a \right) \\ -8\alpha e g^{ik}g^{jl} \left( \partial_i \rho_j^a - \partial_j \rho_i^a \right) \left( \epsilon^{abc} \partial_k\varphi_b \partial_l\varphi_c + \partial_k\varphi_0 \partial_l\varphi_a - \partial_k\varphi_a \partial_l\varphi_0 \right)
    \end{split}.
\end{equation}

For numerical simulations involving the minimization of the energy functional with respect to variations of the metric, it will be convenient to follow the methodology laid out in \cite{Leask_2023,Leask_2024}.
That is, we define the metric independent integrals
\begin{align}
\label{eq: Metric independent E0}
    V = \, & 2 M_\pi^2 \int_{\mathbb{T}^3} \textup{d}^3x \, (1-\varphi_0), \\
\label{eq: Metric independent E2}
    L_{ij} = \, & \int_{\mathbb{T}^3} \textup{d}^3x \, \left\{ \partial_i\varphi_\mu \partial_j\varphi_\mu + 4M_\rho^2 \rho_i^a \rho_j^a \right\}, \\
\label{eq: Metric independent E4}
    \Omega_{ijkl} = \, & \int_{\mathbb{T}^3} \textup{d}^3x \, \left\{ \frac{1}{2}\left( \partial_i\varphi_\mu \partial_k\varphi_\mu \partial_j\varphi_\nu \partial_l\varphi_\nu - \partial_i\varphi_\mu \partial_l\varphi_\mu \partial_j\varphi_\nu \partial_k\varphi_\nu \right) + 2 \left( \partial_i \rho_j^a - \partial_j \rho_i^a \right) \left( \partial_k \rho_l^a - \partial_l \rho_k^a \right) \right. \nonumber \\
    \, & \left. -8\alpha e \left( \partial_i \rho_j^a - \partial_j \rho_i^a \right) \left( \epsilon^{abc} \partial_k\varphi_b \partial_l\varphi_c + \partial_k\varphi_0 \partial_l\varphi_a - \partial_k\varphi_a \partial_l\varphi_0 \right) \right\}.
\end{align}
In terms of these metric independent integrals, the energy is simply
\begin{equation}
    E = E_0 + E_2 + E_4,
\end{equation}
where
\begin{equation}
    E_0 = \sqrt{g} \, V, \quad E_2 = \sqrt{g} \, g^{ij}L_{ij}, \quad E_4 = \sqrt{g} \,g^{ik}g^{jl}\Omega_{ijkl}.
\end{equation}

To understand the role the metric plays in determining crystalline configurations, let us consider the rate of change of the energy  with respect to varying the domain metric $g$ on $\mathbb{T}^3$.
We will do this using the compact notation introduced above.
The first variation of the energy with respect to the variation $g(s)$ of the metric on $\mathbb{T}^3$ is given by
\begin{equation}
    \left.\frac{\textup{d} E(\varphi, \rho,  g_s)}{\textup{d}s}\right|_{s=0} = \int_{\mathbb{T}^3} \textup{d}^3x \sqrt{g}  \braket{S(\varphi, \rho, g), \delta  g}_g,
\end{equation}
where $S(\varphi,\rho, g) \in \Gamma(\odot^2 T^*\mathbb{T}^3)$ is a symmetric $2$-covariant tensor field on $\mathbb{T}^3$, known as the \textit{stress-energy tensor}, defined by
\begin{align}
    \int_{\mathbb{T}^3} \textup{d}^3x \sqrt{g} \, S^{ij} = \, & \frac{1}{2}  g^{ij} \sqrt{g} V + \sqrt{g} \left( \frac{1}{2} g^{ij} g^{kl} -  g^{ik} g^{jl} \right) L_{kl} + \sqrt{g} g^{km} \left( \frac{1}{2} g^{ln} g^{ij}  - 2  g^{il} g^{jn} \right) \Omega_{klmn}.
\end{align}

The space of allowed variations $\mathscr{E}$ is a $6$-dimensional subspace of the space of sections of the rank $6$ vector bundle $\odot^2 T^*\mathbb{T}^3$,
\begin{equation}
    \mathscr{E} = \left\{ \delta g_{ij} \textup{d}x^i \textup{d}x^j \in  \Gamma(\odot^2 T^*\mathbb{T}^3) : \delta g_{ij} \, \textup{constant}, \delta g_{ji} = \delta g_{ij} \right\}.
\end{equation}
By definition, the energy $E$ is critical with respect to variations $g_s$ of the metric if and only if
\begin{equation}
    \left.\frac{\textup{d} E(\varphi,\rho, g_s)}{\textup{d}s}\right|_{s=0} = \int_{\mathbb{T}^3} \textup{d}^3x \sqrt{g} \braket{S(\varphi,\rho, g),\delta g}_g = 0,
\end{equation}
that is, if and only if $S \perp_{L^2} \mathscr{E}$.
Now let the orthogonal compliment of $g$ in $\mathscr{E}$, the space of traceless parallel symmetric bilinear forms, given by
\begin{equation}
    \mathscr{E}_0 = \left\{ \theta \in  \Gamma(\odot^2 T^*\mathbb{T}^3) : \Tr_ g(\theta) = \braket{\theta, g}_ g = 0 \right\}.
\end{equation}
Then the criticality condition $S \perp_{L^2} \mathscr{E}$ can be reformulated as
\begin{equation}
    \int_{\mathbb{T}^3} \textup{d}^3x \sqrt{g} \braket{S(\varphi,\rho, g),  g}_g = 0 \quad \textup{and} \quad S \perp_{L^2} \mathscr{E}_0.
\end{equation}
The first condition $S \perp_{L^2}  g$ is analogous to a virial constraint and the second condition $S \perp_{L^2} \mathscr{E}_0$ coincides with the extended virial constraints derived by Manton \cite{Manton_2009}.
One finds that the condition $S \perp_{L^2} g$ establishes the ``usual'' virial constraint
\begin{equation}
    E_2 - E_4 + 3 E_0 = 0.
\end{equation}

To determine the extended virial constraint corresponding to the condition $S \perp_{L^2} \mathscr{E}_0$, we define a symmetric bilinear form $\Delta: T_x \mathbb{T}^3 \times T_x \mathbb{T}^3 \rightarrow \mathbb{R}$ corresponding to the trace-free part of the stress-energy tensor:
\begin{equation}
    \Delta_{ij} = \sqrt{g} \, L_{ij} + 2 \sqrt{g} \,  g^{kl} \Omega_{ikjl}.
\end{equation}
Then $S \perp_{L^2} \mathscr{E}_0$ if and only if $\Delta$ is orthogonal to $\mathscr{E}_0$ with respect to the inner product $\braket{\cdot,\cdot}_{\mathscr{E}}$.
Therefore, for $\lambda \in \mathbb{R}$ we must have
\begin{equation}
    \Delta = \lambda g.
\end{equation}
Taking the trace of both sides yields
\begin{align}
    3 \lambda = \sqrt{g} \,  g^{ij} L_{ij} + 2 \sqrt{g} \,  g^{ij} g^{kl} \Omega_{ikjl} = E_2 + 2E_4.
\end{align}
Thus, the condition $S \perp_{L^2} \mathscr{E}_0$ produces the extended virial constraint
\begin{equation}
    \Delta = \frac{1}{3} \left( E_2 + 2E_4 \right) g.
\end{equation}
These extended virial constraints act as consistency check on our numerical algorithm, which we detail below.


\subsection{Determining the energy minimizing period lattice}
\label{subsec: Numerical optimization of the lattice geometry}

The general methodology for determining the period lattice that minimizes the static energy was laid out in \cite{Leask_2023}.
The method is identical under redefinition of the metric independent integrals to include the $\rho$-mesons.
We summarize the method here.

Let us fix the Skyrme field $\varphi: \mathbb{T}^3 \rightarrow \SU(2)$ and $\rho$-meson fields $\rho^a_i: \mathbb{T}^3 \rightarrow \mathbb{R}$ and think of the energy as a function of the metric $g$ on $\mathbb{T}^3$, $E_{\varphi,\rho}: \SPD_3 \rightarrow \mathbb{R}$, where $\SPD_3$ is the space of symmetric positive-definite $3 \times 3$-matrices.
To minimize the energy functional with respect to variations of the metric, we use arrested Newton flow on $\SPD_3$.
The essence of the algorithm is as follows: we solve Newton's equations of motion for a particle on $\SPD_3$ with potential energy $E_{\varphi,\rho}$.
Let $g_s$ be a smooth one-parameter curve in $\SPD_3$ with $g_0=F^*d$.
Explicitly, we are solving the system of 2nd order ODEs
\begin{equation}
    \left.\frac{\textup{d}^2}{\textup{d}s^2}\right|_{s=0} (g_{ij})_s = -\frac{\partial E_\varphi}{\partial  g_{ij}} = - \int_{\mathbb{T}^3} \textup{d}^3x \sqrt{g} \, S_{\varphi,\rho}^{ij},
\label{eq: Metric ANF}
\end{equation}
with initial metric $(g_{ij})_0=\vec{X}_i \cdot \vec{X}_j$ and initial velocity $\left.\partial_{s} g_s\right|_{s=0}=0$, where $S_{\varphi,\rho}=S(g)$ is the stress-energy tensor for fixed field configurations $\varphi, \rho$.
Setting $\delta g_s = \partial_s g_s$ as the velocity with $\delta g = \delta g_0$ reduces the problem to a coupled system of first order ODEs.
We implement a fourth order Runge--Kutta method to solve this coupled system.
After each time step $t \mapsto t + \delta t$, we check to see if the energy is increasing.
If $E(t + \delta t) > E(t)$, we take out all the kinetic energy in the system by setting $\delta g(t + \delta t)=0$ and restart the flow.
The flow then terminates when every component of the stress-energy tensor $S_{\varphi,\rho}$ is zero to within a given tolerance (we have used $10^{-6}$).


\subsection{Crystalline configurations}
\label{subsec: Crystalline configurations}

With the numerical method in place, we now turn our focus to constructing $\rho$-skyrmion crystals.
The starting point in the construction of $\rho$-skyrmion crystals is the Fourier series-like expansion of the Skyrme fields as an initial configuration \cite{Kugler_1989},
\begin{subequations}
    \begin{align}
        \sigma = \, & \sum_{a,b,c}^{\infty} \beta_{abc} \cos\left(\frac{2a\pi x}{L}\right) \cos\left(\frac{2b\pi y}{L}\right) \cos\left(\frac{2c\pi z}{L}\right), \\
        \pi_1 = \, & \sum_{h,k,l}^{\infty} \alpha_{hkl} \sin\left(\frac{2h\pi x}{L}\right) \cos\left(\frac{2k\pi y}{L}\right) \cos\left(\frac{2l\pi z}{L}\right), \\
        \pi_2 = \, & \sum_{h,k,l}^{\infty} \alpha_{hkl} \cos\left(\frac{2l\pi x}{L}\right) \sin\left(\frac{2h\pi y}{L}\right) \cos\left(\frac{2k\pi z}{L}\right), \\
        \pi_3 = \, & \sum_{h,k,l}^{\infty} \alpha_{hkl} \cos\left(\frac{2k\pi x}{L}\right) \cos\left(\frac{2l\pi y}{L}\right) \sin\left(\frac{2h\pi z}{L}\right),
    \end{align}
\label{eq: Skyrme crystal Fourier series}
\end{subequations}
with initial metric $g=L^3 \Id_3$ on $\mathbb{T}^3$.
For the $\textup{SC}_{1/2}$ crystal, the Fourier coefficients $\beta_{abc}$ and $\alpha_{hkl}$ must satisfy the conditions: $h$ and $k$ are odd, and $l$ is even, and $a,b,c$ are all odd \cite{Huidobro_2023}.
Then, truncating the Fourier series \eqref{eq: Skyrme crystal Fourier series} to only include the first terms in the expansions yields the approximation of Castillejo \textit{et al.} \cite{Castillejo_1989}.

From the $\textup{SC}_{1/2}$ crystal, the other three crystals can be constructed by applying a chiral $\SO(4)$ transformation $Q\in\SO(4)$, such that $\varphi=Q\varphi_{1/2}$, and minimizing the energy with respect to variations of the field and the lattice.
These chiral transformations $Q\in\SO(4)$ can be determined by considering a deformed energy functional on the moduli space of critical points of the Skyrme energy functional, and are found to be
\begin{equation}
    Q \in \left\{ \Id_4,
    \underbrace{\begin{pmatrix}
        (0,1,1,1)/\sqrt{3} \\
        * \\
    \end{pmatrix}}_{Q_\alpha},
    \underbrace{\begin{pmatrix}
        (0,0,0,1) \\
        * \\
    \end{pmatrix}}_{Q_\textup{multiwall}},
    \underbrace{\begin{pmatrix}
        (0,0,1,1)/\sqrt{2} \\
        * \\
    \end{pmatrix}}_{Q_\textup{chain}}
    \right\}.
\label{eq: Chiral Q transformations}
\end{equation}
The other three rows of the chiral transformations $Q_\alpha$, $Q_\textup{multiwall}$ and $Q_\textup{chain}$, labeled by the asterisk, can be obtained by using the Gram--Schmidt process.

\begin{figure}[t]
    \centering
    \begin{subfigure}[b]{0.45\textwidth}
        \includegraphics[width=\textwidth]{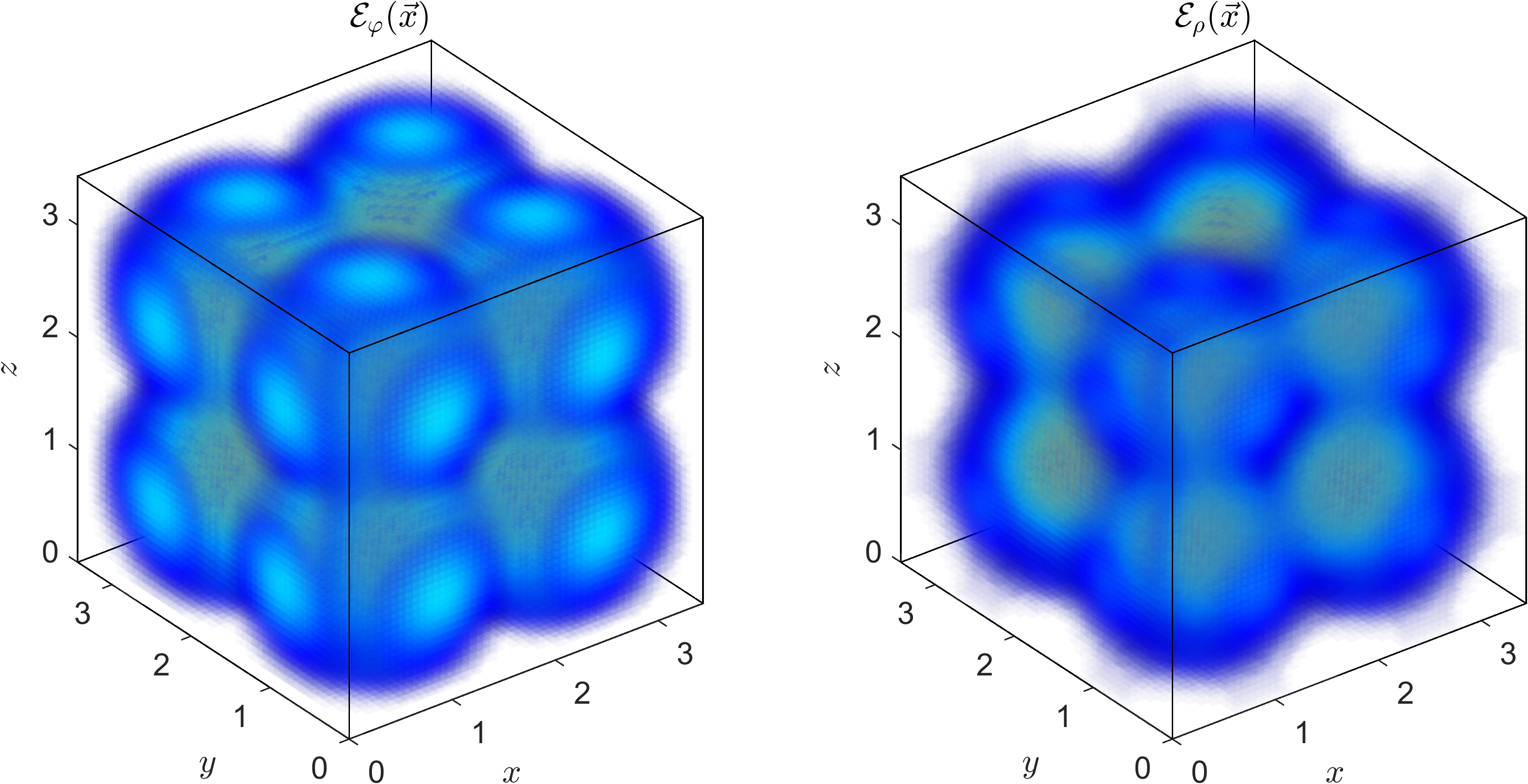}
        \caption{$c_\alpha=0.125$}
        \label{fig: Alpha crystal low alpha}
    \end{subfigure}
    ~
    \begin{subfigure}[b]{0.45\textwidth}
        \includegraphics[width=\textwidth]{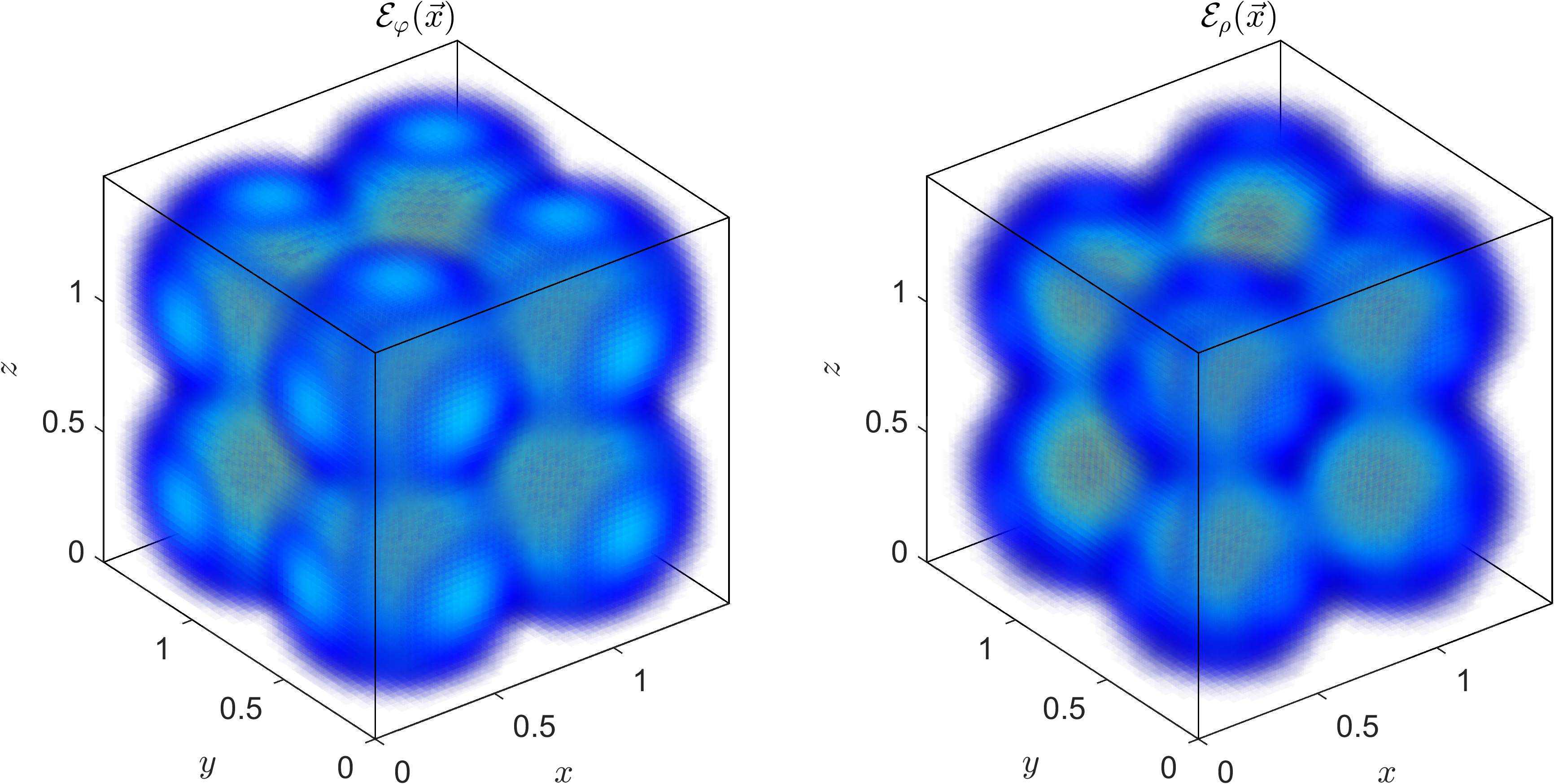}
        \caption{$c_\alpha=0.225$}
        \label{fig: Alpha crystal low alpha}
    \end{subfigure}
    \caption{Energy density plots of the Skyrme $\mathcal{E}_\varphi$ and $\rho$-meson $\mathcal{E}_\rho$ contributions to the $\alpha$-particle crystal for (a) low and (b) high coupling constants. At higher couplings, the crystals are more lightly bound and support $\alpha$-particle clustering.}
    \label{fig: alpha crystals}
\end{figure}

Following the methodology laid out in this paper for determining crystals in the $\rho$-Skyrme model, with the initial configuration as detailed above, we find that {\it the ground state configuration in this theory is the $\alpha$-particle crystal with cubic lattice geometry}.
The second lowest energy configuration is found to be the multiwall crystal.
As the value of the coupling constant increases, the energy difference between these two solutions becomes greater and more apparent.
Therefore, we will use the $\alpha$-particle crystal to determine the compression modulus in this theory.
Energy density plots of the Skyrme field $\varphi$ and $\rho$-meson contributions to the $\alpha$-particle crystal, for low and high $c_\alpha$ couplings, are shown in Fig.~\ref{fig: alpha crystals}.

\section{Resolving the compression modulus problem}
\label{sec: Resolving the compression modulus problem}

\begin{figure}
    \centering
    \includegraphics[width=0.6\columnwidth]{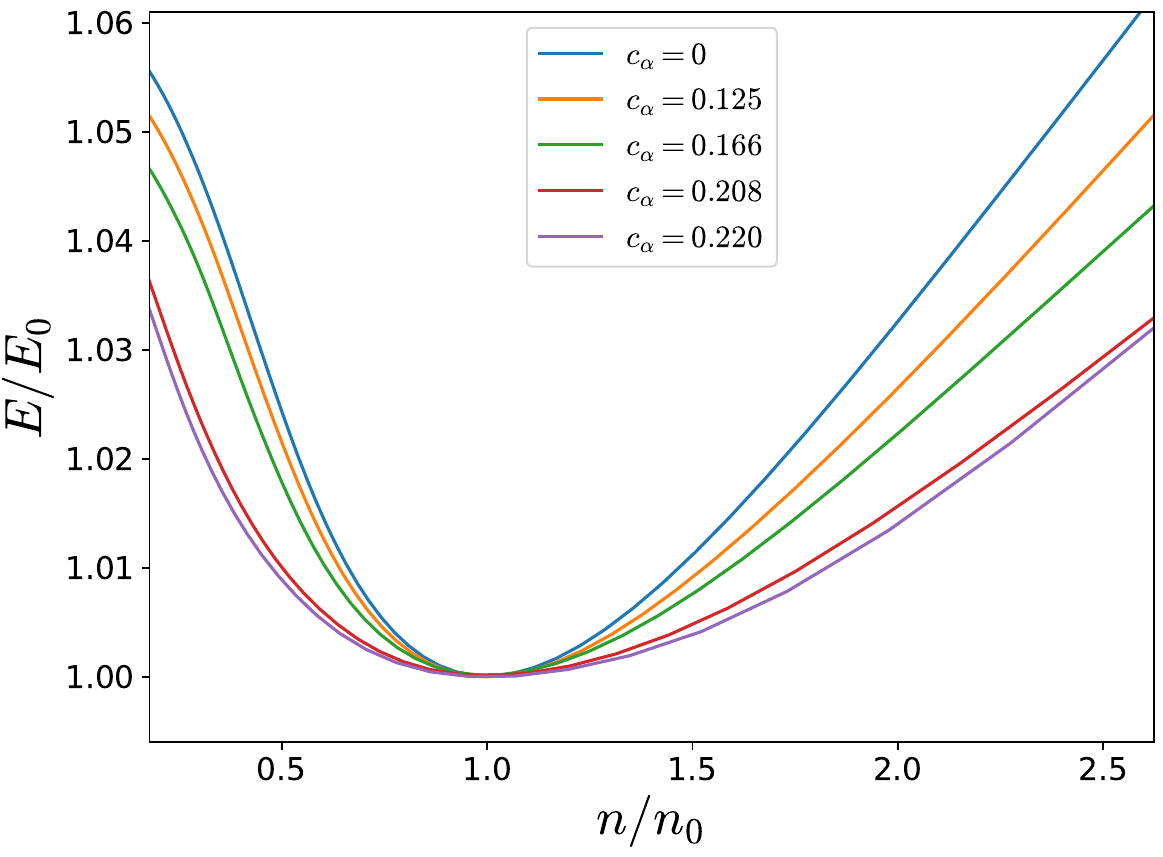}
    \caption{Comparison of the equations of state of the $\alpha$-particle crystal about saturation for various choices of the $\rho$-meson coupling constant $c_\alpha=\alpha e$. It can be clearly seen that the compression modulus decreases as the coupling is increased.} 
    \label{fig: Compression modulus}
\end{figure}

To compute the compression modulus within the $\rho$-Skyrme model, we need to construct skyrmion crystals at various nuclear densities $n_B$.
This is achieved by constraining the fundamental unit cell of the period lattice $\Lambda$ to have fixed volume $V$.
This amounts to setting $\det(g)$ constant and, in turn, fixes the baryon density $n_B = B/V$.
In practice, we replace the stress-energy tensor $S_{ij}$ by its projection $S_{ij}-\frac{1}{3}S_{kl}g^{kl}g_{ij}$.
As the ground state is observed to be the $\alpha$-particle crystal, we use this crystal to compute the compression modulus.
This is carried out at various values of the coupling constant $c_\alpha$ and the resulting equations of state about saturation are shown in Fig.~\ref{fig: Compression modulus}.

It can be clearly seen that the equations of state about saturation get more shallow as the coupling constant $c_\alpha$ is increased.
Consequently, this means that the compression modulus is getting smaller in value.
Furthermore, the binding energies also decrease as the coupling $c_\alpha$ increases.
This is summarized in Tab.~\ref{tab: Compression modulus}.

We now provide a simple argument that shows why a theory with low binding energies will also produce a lower value of the compression modulus.
If the binding energies decrease then the asymptotic energy (as $n_B \rightarrow 0$) of the crystalline solution must also decrease.
This reduces the energy difference between the asymptotic crystal (in our case, the isolated $\alpha$-particle) and the ($\alpha$-particle) crystal at saturation, which naturally reduces the curvature of the equation of state at saturation.
Hence, a lower value of the compression modulus.

\begin{table}[t]
    \centering
    \begin{tabular}{|c|c|c|c|}
        \hline
        $c_\alpha$ & $K_0/E_0$ & $K_0$ (MeV) & BE ($\%$) \\
        \hline
        $0$ & $1.170$ & $1080$ & $5.54$ \\
        $0.125$ & $0.985$ & $909$ & $5.36$ \\
        $0.166$ & $0.778$ & $718$ & $5.00$ \\
        $0.208$ & $0.461$ & $425$ & $4.25$ \\
        $0.220$ & $0.381$ & $351$ & $3.85$ \\
        \hline
    \end{tabular}
    \caption{Values of the adimensional coupling constant between $\rho$-mesons and pions with the corresponding values of the compression modulus and energy ratio. In the last two columns we compute the difference between the minimum and the $n_B \rightarrow 0$ asymptotic energy and also the corresponding value of the physical coupling constant, respectively.}
    \label{tab: Compression modulus}
\end{table}


\section{Concluding remarks}
\label{sec:conclusion}

In the present paper we studied skyrmion crystals in an extended version of the Skyrme model where, the mesonic degrees of freedom are assisted by the $\rho$ vector meson.
This is a natural extension of the original Skyrme's idea, both from a phenomenological as well as a theoretical point of view.
The $\rho$-meson is the next lightest mesonic particle.
It is also naturally connected with the $\SU(2)$ pionic field via the dimensional reduction of (4+0) Yang-Mills theory \`a la Sutcliffe. 

Specifically, we considered a minimal extension with only one interaction term, which gives rise to the usual two-pion vertex.
Already at this level we obtained quite non-trivial and physically very encouraging results. 

First of all, we found that the ground state is the $\alpha$-particle crystal.
This solution becomes more and more favored as the $\rho$-meson coupling constant grows.
This should be contrasted with the generalized Skyrme model where the multiwall solution forms the ground state for a large range of parameters. It also differs from the $\omega$-Skyrme crystal where the geometry of the ground state strongly depends on the values of the coupling constants. 

Secondly, which is the most important result of the current work, the inclusion of $\rho$-meson significantly reduces the value of the compression modulus.
For $c_\alpha = 0.220$, which is the $\rho$-meson coupling constant close to the maximal value, $K_0$ decreases from original 1080 MeV to 351 MeV.
This is a striking improvement.
Apparently, the $\rho$-meson makes the equation of state much softer at the saturation point which is a very welcome property of the considered model. 

An obvious question is whether the equation of state is stiff enough at higher densities to support neutron stars with masses in agreement with the current astrophysical data.
This may require an addition of the sextic term and/or inclusion of the $\omega$-meson.
In fact, this meson is only 12 MeV heavier than $\rho$-meson.
A variant of the Skyrme model with $\omega$-meson has been very recently considered.
Similarly, a significant improvement of the compression modulus has been reported \cite{Leask_Harland_2024}. 

Analytic methods have also been employed to study vector meson extensions of the Skyrme model \cite{Barriga:2023jam,Barriga_2022}.
It has also been natural to consider generalizations of the non-linear pion theory by replacing the \textit{ad hoc} Skyrme term with explicit interactions with vector mesons.
One such approach is based on the so-called hidden local symmetry (HLS) method, where a hidden symmetry of the NL$\sigma$M is gauged and the corresponding gauge particle acquires mass through the Higgs mechanism \cite{Forkel_1991}.
This allows for the incorporation of $\rho$-mesons, as well as the $\omega$-meson.

The HLS approach has been investigated extensively, especially in the context of dense nuclear matter \cite{Ma_2012,Yong-Liang_2013,Harada_2013,Harada_2014,Paeng_2016}.
Within the HLS framework, Lee \textit{et al.} \cite{Park_2003} incorporated a dilaton field to the massive Skyrme model, which is associated with the scale anomaly of QCD.
In their model the dilaton is crucial in realizing the phase transition from the Goldstone mode (spontaneously broken chiral symmetry $\textup{FCC}_{1}$ phase) to the Wigner mode (unbroken chiral symmetry $\textup{SC}_{1/2}$ phase) consistently with the vector manifestation (VM) fixed point.
This fixed point is characterized by the vanishing of both $\braket{\sigma}$ and the in-medium pion decay constant $F_{\pi}^*$, corresponding to the restoration of the spontaneously broken chiral symmetry.
Interestingly, the addition of the $\omega$-meson prevents the scale-anomaly dilaton field, and thus the in-medium pion decay constant, from developing a vanishing vacuum expectation value at the VM fixed point \cite{Park_2004}, resulting in a non-restoration of chiral symmetry.
This is reminiscent of pseudo-gap phenomena in condensed matter physics.

In all of the above HLS crystal investigations, the Kugler and Shtrikmann Fourier series method was generalized to incorporate vector mesons.
However, as was observed in this paper and in \cite{Leask_Harland_2024}, the $\textup{SC}_{1/2}$ crystal is in fact not the ground state configuration in the Skyrme model with vector mesons.
Therefore, generalizing the method presented in \cite{Leask_2023} to the Skyrme model with both, $\rho$ and $\omega$, mesons included seems to be a natural next step of the investigation. 


\section*{Acknowledgments}
A. W. and C. N. were supported by the Polish National Science center (NCN 2020/39/B/ST2/01553).
P.~L. acknowledges funding from UKRI, Grant No. EP/V520081/1.


\appendix


\section{Derrick scaling of the $\mathcal{L}_{24}$-Skyrme model}
\label{derrick}

Consider a variation $\varphi_\lambda: \mathbb{R}^3 \times \mathbb{R} \rightarrow S^3$ of the Skyrme field $\varphi$ such that $\varphi_{\lambda=0}=\varphi$.
This has infinitesimal generator $\partial_\lambda \varphi_\lambda|_{\lambda=0} \in \Gamma(\varphi^{-1}T S^3)$, where $\varphi^{-1}T S^3$ is the vector bundle  over $\mathbb{R}^3$ with fibre $T_{\varphi(x)}S^3$ over $x\in \mathbb{R}^3$.
Explicitly, if we consider the spatial rescaling $x \mapsto e^\lambda x$, then we have a one-parameter family of maps $\varphi_\lambda=\varphi(e^\lambda x)$ such that $\varphi_{\lambda=0}=\varphi$.
The rescaled massless $\mathcal{L}_{24}$ static energy functional is then
\begin{equation}
    E_\lambda = E_{24}[\varphi_\lambda]  = e^{\lambda} E_2 + e^{-\lambda} E_4.
\label{eq: Skyrme model - Derricks theorem}
\end{equation}
If the Skyrme field configuration $\varphi$ is a minimiser of the $\mathcal{L}_{24}$-energy $E$, then we require
\begin{equation}
    \left.\frac{\textup{d}}{\textup{d}\lambda}\right|_{\lambda=0} E_{24}[\varphi_\lambda] = E_2 - E_4 = 0,
\end{equation}
which yields the familiar virial constraint $E_2=E_4$.

The true $\mathcal{L}_{24}$-energy minimizing crystal is the cubic lattice of half-skyrmions found independently by Kugler \& Shtrikmann \cite{Kugler_1988} and Castillejo \textit{et al.} \cite{Castillejo_1989}.
Let us denote this minimal energy crystalline solution of the $\mathcal{L}_{24}$-Skyrme model by $\varphi_0 \equiv \varphi(L_0)$, where $L_0$ corresponds to the side length of the energy minimizing cubic lattice.
Under the one-parameter variation $L_0 \mapsto e^{\lambda}L_0$, the rescaled half-crystal configuration $\varphi(e^{\lambda}L_0)$ approximates the true minimizer at volume $V=e^{3\lambda}L_0^3$ well for small $\lambda$.
Then, the compression modulus may be related to the second derivative of the energy with respect to the scaling factor $\lambda$,
\begin{align}
    K_0 = \left.\frac{\textup{d}^2}{\textup{d}\lambda^2} \right|_{\lambda=0} E_{24}[\varphi(e^{\lambda}L_0)]  = E_2[\varphi_0] + E_4[\varphi_0]  = E_0.
\end{align}
Thus, one finds that $K_0/E_0=1$ and, hence, this leads to a value for the compression modulus that is roughly four times too large.

\bibliography{bib.bib}

\end{document}